\documentclass[12pt]{article}
\usepackage{epsfig}
\usepackage{a4}
\usepackage{latexsym}
\usepackage{cite}
\usepackage{slashed}
\usepackage{hyperref}
\usepackage{tikz}
\usepackage{axodraw2}

\usepackage{color,colordvi}

\newcommand{\gsim}{\raisebox{-0.07cm}{$\:\:\stackrel{>}{{\scriptstyle
 \sim}}\:\: $} }

\newcommand{\equal}{\:\: = \:\:}

\newcommand{\hspn}{{\hspace{-5mm}}}
\newcommand{\hspp}{{\hspace{6mm}}}
\newcommand{\beq}{\begin{equation}}
\newcommand{\eeq}{\end{equation}}
\newcommand{\bea}{\begin{eqnarray}}
\newcommand{\eea}{\end{eqnarray}}
\newcommand{\nn}{\nonumber}
\newcommand{\MSb}{$\overline{\mbox{MS}}$}
\newcommand{\ra}{\rightarrow}

\newcommand{\alsMM}{\alpha_{\rm s,MM}}
\newcommand{\als}{\alpha_{\rm s}}
\newcommand{\ars}{a_{\rm s}}
\newcommand{\ep}{\varepsilon}

\begin{document}
\setlength{\parskip}{0.2cm}
\setlength{\baselineskip}{0.535cm}

\def\z#1{{\zeta_{#1}}}
\def\zss{\zeta_2^{\:\!2}}
\def\zst{\zeta_2^{\:\!3}}
\def\zts{\zeta_3^{\:\!2}}

\def\nc{{n_c}}
\def\ncs{{n_{c}^{\,2}}}
\def\nct{{n_{c}^{\,3}}}

\def\ca{{C^{}_{\!A}}}
\def\cas{{C^{\,2}_{\!A}}}
\def\cat{{C^{\,3}_{\!A}}}
\def\caf{{C^{\,4}_{\!A}}}
\def\cai{{C^{\,5}_{\!A}}}

\def\cf{{C^{}_F}}
\def\cfs{{C^{\, 2}_F}}
\def\cft{{C^{\, 3}_F}}
\def\cff{{C^{\, 4}_F}}

\def\nf{{n^{}_{\! f}}}
\def\nfz{{n^{\:\!0}_{\! f}}}
\def\nfo{{n^{\:\!1}_{\! f}}}
\def\nfs{{n^{\:\!2}_{\! f}}}
\def\nft{{n^{\:\!3}_{\! f}}}
\def\nff{{n^{\:\!4}_{\! f}}}

\def\nl{{n^{}_{\! f}}}
\def\nlz{{n^{\:\!0}_{\! f}}}
\def\nlo{{n^{\:\!1}_{\! f}}}
\def\nls{{n^{\:\!2}_{\! f}}}
\def\nlt{{n^{\:\!3}_{\! f}}}
\def\nlf{{n^{\:\!4}_{\! f}}}

\def\tf{{\:\!T^{}_{\!F}}}
\def\tfs{{\:\!T^{\,2}_{\!F}}}
\def\tft{{\:\!T^{\,3}_{\!F}}}
\def\tff{{\:\!T^{\,4}_{\!F}}}

\def\dfAAna{{\frac{d_A^{\,abcd}d_A^{\,abcd}}{N_A }}} 
\def\dfFAna{{\frac{d_F^{\,abcd}d_A^{\,abcd}}{N_A }}}
\def\dfFFna{{\frac{d_F^{\,abcd}d_F^{\,abcd}}{N_A }}}
\def\dfFAnr{{\frac{d_F^{\,abcd}d_A^{\,abcd}}{N_R }}}
\def\dfFFnr{{\frac{d_F^{\,abcd}d_F^{\,abcd}}{N_R }}}
\def\dtFFnr{{\frac{d_F^{\,abc}d_F^{\,abc}}{N_R }}}

\def\as(#1){{\alpha_{\rm s}^{\,#1}}}
\def\ar(#1){{a_{\rm s}^{\,#1}}}

\def\MZ{{M_{\rm Z}}}
\def\MZs{{M_{\rm Z}^{\:\!2}}}

\def\MHt{{M_{\rm H}^{\:\!3}}}
\def\MHs{{M_{\rm H}^{\:\!2}}}
\def\MH{{M_{\rm H}}}

\def\Lt(#1){{L_t^{\:\!#1}}}
\def\lntwo(#1){{\ln^{\:\!#1\!}2}}

\def\g#1{{g_{#1}^{}}}
\def\gam#1{{\gamma_{#1}^{}}}
\def\Gam(#1,#2){{\gamma_{#1}^{\,#2}}}
\def\rs#1{{\tilde{r}_#1^{}}}

\def\b#1{{\beta_{#1}}}
\def\B(#1,#2){{\beta_{#1}^{\,#2}}}

\def\muRs{{\mu_R^{\,2}}}
\def\L{\mathcal{L}}
\def\eps{\epsilon}
\def\dots{..}

\def\frct#1#2{\mbox{\large{$\frac{#1}{#2}$}}}



\begin{titlepage}
\noindent
Nikhef 2017-029 \hfill {\tt arXiv:1707.01044v2}\\
LTH 1136 \\
\vspace{1.4cm}
\begin{center}
{\Large \bf On Higgs decays to hadrons and the R-ratio at N$^{\bf 4}$LO}\\
\vspace{3cm}
\large
F. Herzog$^{\:\!a}$, B. Ruijl$^{\:\! a,b}$, T. Ueda$^{\:\! a}$, 
J.A.M. Vermaseren$^{\:\! a}$ and A. Vogt$^{\:\! c}$\\
\vspace{1.2cm}
\normalsize
{\it $^a$Nikhef Theory Group \\
\vspace{0.1cm}
Science Park 105, 1098 XG Amsterdam, The Netherlands} \\
\vspace{0.5cm}
{\it $^b$Leiden Centre of Data Science, Leiden University \\
\vspace{0.1cm}
Niels Bohrweg 1, 2333 CA Leiden, The Netherlands}\\
\vspace{0.5cm}
{\it $^c$Department of Mathematical Sciences, University of Liverpool\\
\vspace{0.1cm}
Liverpool L69 3BX, United Kingdom}\\
\vspace{2.5cm}
{\large \bf Abstract}
\vspace{-0.2cm}
\end{center}
We present the first determination of Higgs-boson decay to hadrons at the 
next-to-next-to-next-to-next-to-leading order of perturbative QCD in the limit 
of a heavy top quark and  massless light flavours.
This result has been obtained by computing the absorptive parts of the relevant
five-loop self-energy for a general gauge group and combining the outcome with
the corresponding coefficient function already known to this order in~QCD. 
Our new result reduces the uncertainty due to the truncation of the 
perturbation series to a fraction of the uncertainty due to the present error 
of the strong coupling constant.
We have also performed the corresponding but technically simpler computations 
for direct Higgs decay to bottom quarks and for the electromagnetic $R$-ratio 
in $e^+e^- \ra \mbox{ hadrons}$, thus verifying important fifth-order results 
obtained only by one group so far.

\vspace*{0.5cm}
\end{titlepage}

%
\section{Introduction}
\label{sec:introduction}

The production and decay processes of the Higgs boson, discovered five 
years ago at CERN \cite{HiggsA,HiggsC} with a mass $\MH$ of 125 GeV, are 
among the most important research topics in collider physics. The dominant 
standard-model decay is that to bottom quarks, $H \ra \bar{b}^{}b$
(+$\,$hadrons).
The QCD calculations of this decay mode have been completed up to the 
fourth order in the strong coupling $\als$, see ref.~\cite{DaviesSW17} 
and references therein.
A crucial component of this high accuracy is the next-to-next-to-%
next-to-next-to-leading order (N$^4$LO) computation \cite{HbbN4LO} of the
decay to quarks via their direct coupling to the Higgs. This calculation,
in which the quark mass can be neglected (except in the Yukawa coupling), 
has not been repeated so~far.

The second important hadronic decay channel arises via $H \ra gg$, where
the coupling of the Higgs to gluons is predominantly mediated, in the
standard model, by the top quark. Due to $\MH \ll 2^{}m_{\:\!t}$,
high-order QCD corrections to this process can be evaluated in an
effective theory in which the top quark has been integrated out
\cite{Inami:1982xt}; for 1/$m_{\:\!t}$ corrections up to NNLO see
refs.~\cite{Larin:1995sq,Schreck:2007um}.
The resulting coefficient function for the effective Higgs coupling to
gluons is known to N$^4$LO
\cite{KLS96,dec-as3,dec-as4a,dec-as4b,Kniehl:2006bg,ChBK-LL16}.
The absorptive part of the corresponding vacuum polarization is not yet
known at this order. The N$^3$LO corrections have been computed in ref.~%
\cite{HggN3LO} and checked 
in ref.~\cite{MV2} (except for their kinematic $\pi^2$ terms)
and very recently in ref.\ \cite{Davies:2017rle};
see refs.\ \cite{Inami:1982xt,Djouadi:1991tka,HggNNLO} for the previous orders.

In this article we present this hitherto missing fourth-order correction, 
thus completing the N$^4$LO corrections to Higgs decay into hadrons in 
the limit of a heavy top quark and any number of massless flavours.
We have also performed the computationally far simpler N$^4$LO determination 
of the $H \rightarrow \bar{b}^{}b$ decay rate and verified the result of 
ref.~\cite{HbbN4LO}.
A somewhat more demanding but closely related computation is that of the
N$^4$LO corrections to the electromagnetic $R$-ratio for the process
$e^+e^- \ra \mbox{ hadrons}$. So far these corrections were determined 
only by one group \cite{Rratio1,Rratio2,Rratio3,RratioRC}. 
We have also re-calculated this quantity to the fourth order, and find 
complete agreement for both the non-singlet and singlet contributions.

Our N$^4$LO computations employ the same overall strategy as those by the 
Karlsruhe-Moscow group mentioned above (see also ref.~\cite{BchK15rev}):
the pole terms of the relevant correlation function are calculated in 
dimensional regularization \cite{DimReg1,DimReg2} at five loops, and 
subsequently the absorptive part is extracted.
Our use of this approach has been made possible by the development of
(a)
{\sc Forcer} \cite{tuACAT2016,tuLL2016,FORCER}, a~{\sc Form}
\cite{FORM3,TFORM,FORM4} program for the parametric reduction of four-loop
self-energy integrals, and
(b)
a program \cite{HerzogRuijl17} efficiently implementing the $R^*$~operation, 
see refs.~\cite{RSTAR82,RSTAR84,Batkovich1,Batkovich2,RSTAR91}, locally
for the evaluation of $L\,$-loop pole terms in terms of \mbox{%
$(L-\!1)\:\!$-$\:\!$loop} integrals. In order to cope with the computations
for $H \rightarrow gg$, which are far more demanding than those required to
determine the five-loop beta function \cite{beta4a,beta4b}, the latter
program has undergone substantial modifications and extensions.

The remainder of this article is organized as follows: 
in section~\ref{sec:framework} we define our notations and briefly address 
some computational details.  
Our new N$^4$LO result for the $H \!\ra\! gg$ decay width is presented 
and discussed in section \ref{sec:hgg}. Due to the rather large higher-order
coefficients in the expansion in $\als$, it is interesting to compare the
results in the standard \MSb\ scheme \cite{MS,MSbar} to those in a fairly 
common (and, in certain contexts, more physical) alternative, the miniMOM 
scheme \cite{MiniMOM1,MiniMOM2}.
The transformation to this scheme, in contrast to other MOM schemes, is known 
to N$^4$LO \cite{N4LOmMOM}; it has argued to be preferable to \MSb\ for
$H \ra gg$ in a recent N$^3$LO study \cite{Zeng:2015gha}.

In section \ref{sec:hbb} we briefly address the decay $H \ra \bar{b}^{}b$. 
We present the N$^4$LO correction for a general gauge group and a general 
renormalization scale which has not been written down in the literature 
so far.
The N$^4$LO results in QCD, which show a far less problematic behaviour at the
only relevant case of $\als \approx 0.1$ than their $H \ra gg$ counterparts,
have been known and discussed for more than ten years. Hence there is no
need to go into more detail in this case. This is somewhat different for the
$R$-ratio addressed in section \ref{sec:rratio}, despite its even smaller
coefficients in the expansion in $\als$, since this quantity is of physical
relevance down to rather low scales and correspondingly high valus of $\als$.
Hence the size and scale (in-)$\,$stabilily of this quantity is illustrated in 
both the \MSb\ and the miniMOM scheme at two low-scale reference points.
We briefly summarize our results in section \ref{sec:summary}.

%
\medskip
\section{Theoretical framework and calculations}
\label{sec:framework}
\setcounter{equation}{0}
 
\vspace*{-1mm}
\subsection*{Inclusive Higgs-boson decay to gluons}
\vspace*{-1mm}
In the limit of a large top-quark mass and $\nl$ effectively massless 
flavours, the decay of the Higgs boson to hadrons can be calculated using 
the effective Lagrangian \cite{Inami:1982xt,HggNNLO}
\beq
\label{eq:Leff}
  \L_{\mathrm{eff}} \;=\; \L_{\mathrm{QCD}(\nl)} \,-\, 2^{1/4\,}
  G_{\rm F}^{\:\!1/2\,} C_1 \:\!H \:\! G^{\:\!\mu\nu}_a G_{\mu\nu}^{\:\!a}
\; .
\eeq
Here $H$ is the Higgs field, and $G^{\:\!\mu\nu}_a$ the renormalized gluon 
field-strength tensor for QCD with $\nl$ flavours and the Lagrangian
$\L_{\mathrm{QCD}(\nl)}$. 
The renormalized coefficient function $C_1$ includes the top-mass dependence. 
$G_{\rm F} \simeq 1.1664\cdot 10^{-5}\, \mbox{GeV}^{-2}$ denotes the Fermi 
constant. 

At the leading order (LO) of perturbative QCD, eq.~(\ref{eq:Leff}) implies 
that the Higgs decays to hadrons only via $H \ra gg$. At the (next-to-)$^n$-%
leading order, N$^n$LO, up to $n$ additional partons occur in the final state.
As usual, we will refer to the inclusive decay induced by eq.~(\ref{eq:Leff}) 
as $H \ra gg$ also beyond LO. 
The corresponding partial decay width $\Gamma_{H \to\,gg}$ can be related, via 
the optical theorem, to the imaginary part of the Higgs-boson self energy: 
\beq
\label{GamHgg}
  \Gamma_{H\to\, gg} \;=\; \frac{\sqrt{2}\, G_{\rm F}}{\MH}\: |C_1|^2 \,
  \mathrm{Im}\, \Pi^{GG}(-\MHs-i\delta)
\; ,
\eeq
where $\MH$ is the Higgs boson mass, $\delta$ is an infinitesimally small positive real parameter and $\Pi^{GG}$ denotes the contribution to 
the self energy of the Higgs boson which is induced by its effective gluonic 
couplings as produced by eq.~(\ref{eq:Leff}).

The Wilson coefficient $C_1$ can be extracted via a low-energy theorem from 
a decoupling relation \cite{dec-as3} which relates the value of the strong 
coupling in a theory with $\nl$ light flavours,
\beq
  \als(\mu^2)  \;\equiv\; \as({(\nl)})(\mu^2)
\; ,
\eeq 
to its value $\as({(\nl+1)})$ in the corresponding theory with $\nl$ light 
flavours and one heavy flavour. 
The analytic QCD expression for $C_1$ up to N${}^4$LO has been provided in
ref.~\cite{ChBK-LL16} as a function of $\as({(\nl+1)})$ at the renormalization 
scale $\mu=\mu_t$, where $\mu_t=m_t(\mu_t)$ is the scale invariant (SI) top 
quark mass, i.e., the \MSb{} mass evaluated at scale $\mu_t$. 
Using the decoupling relation \cite{Kniehl:2006bg,dec-as4a,dec-as4b,ChBK-LL16}, 
the renormalization group and the three-loop relation between the \MSb\ 
mass and the on-shell (OS) mass \cite{Chetyrkin:1999qi,Melnikov:2000qh}, 
we have rewritten the four-loop Wilson coefficient as a function of 
$\as({(\nl)})(\mu^2)$ at an arbitrary renormalization scale $\mu$ for the 
SI, \MSb\ and OS top quark masses. The same has recently been done to three 
loops for the OS scheme in ref.~\cite{DaviesSW17}.

For the convenience of the reader we include the resulting analytic
expressions for $C_1$. These are presented in the form
\beq
\label{C1exp} 
  C_{1,\,\rm X} \;=\; \mbox{}
  - \frct{1}{3}\, \ars \,\Big( \,
  1 + \sum_{n=1} \, c_{n,\,\rm X} \, \ar(n)(\mu^2) \Big)
\quad \mbox{ with } \quad
  \ars \:\equiv\: \frac{\als}{4\:\!\pi}
\:\: .
\eeq
Here $\rm X$ labels the mass scheme employed, and we have indicated the
reduced coupling $\ars$ that we employ for all analytic expressions.
The first two coefficients are the same in the above top-mass schemes up to 
the different definitions of masses entering $\Lt() = \ln (\mu^2/m_t^2)$,
\beq
  c_1^{} \;=\;
         11
\;, \qquad
  c_2^{} \;=\;
            { 2777 \over 18 }
          + 19\, \* \Lt()
       - \nl \, \*  \bigg[
            { 67 \over 6 }
          - { 16 \over 3 }\, \* \Lt()
          \bigg]
\; .
\eeq
The N$^3$LO and N$^4$LO coefficients in the SI scheme read
\bea
  c_{3,\,\rm SI}^{} &\! =\! &
          - { 2892659 \over 648 }
          + { 897943 \over 144 }\, \* \z3
          + { 4834 \over 9 }\, \* \Lt()
          + 209\, \* \Lt(2)
\nn \\[1mm] & & \mbox{\hspn}
       + \nl \, \*  \bigg[ \,
            { 40291 \over 324 }
          - { 110779 \over 216 }\, \* \z3
          + { 2912 \over 27 }\, \* \Lt()
          + 46\, \* \Lt(2)
          \bigg]
\nn \\[1mm] & & \mbox{\hspn}
       - \nls \, \*  \bigg[ \,
            { 6865 \over 486 }
          - { 77 \over 27 }\, \* \Lt()
          + { 32 \over 9 }\, \* \Lt(2)
          \bigg]
 \:\: , \\[3mm]
  c_{4,\,\rm SI}^{} &\! =\! &
          - { 854201072999 \over 2041200 }
          + { 28121193841 \over 75600 }\, \* \z3
          + { 4674213853 \over 28350 }\, \* \zss
          + { 913471669 \over 3780 }\, \* \z5
\nn \\[1mm] & & \mbox{}
          - { 577744954 \over 4725 }\, \* \lntwo()\, \* \zss
          + { 93970579 \over 567 }\, \* \lntwo(2)\, \* \z2
          - { 84531544 \over 2835 }\, \* \lntwo(3)\, \* \z2
\nn \\[1mm] & & \mbox{}
          - { 93970579 \over 3402 }\, \* \lntwo(4)
          + { 42265772 \over 14175 }\, \* \lntwo(5)
          - { 375882316 \over 567 }\, \* a_4^{}
          - { 338126176 \over 945 }\, \* a_5^{}
\nn \\[1mm] & & \mbox{}
          - { 47987641 \over 216 }\, \* \Lt()
          + { 9364157 \over 48 }\, \* \z3 \, \* \Lt()
          + { 29494 \over 3 }\, \* \Lt(2)
          + 2299\, \* \Lt(3)
\nn \\[1mm] & & \mbox{\hspn}
       + \nl \, \*  \bigg[ \,
            { 76094378783 \over 2041200 }
          - { 12171659669 \over 151200 }\, \* \z3
          + { 608462731 \over 113400 }\, \* \zss
          - { 22104149 \over 1890 }\, \* \z5
\nn \\[1mm] & & \mbox{}
          + { 37273868 \over 4725 }\, \* \lntwo()\, \* \zss
          - { 11679301 \over 1134 }\, \* \lntwo(2)\, \* \z2
          + { 5453648 \over 2835 }\, \* \lntwo(3)\, \* \z2
\nn \\[1mm] & & \mbox{}
          + { 11679301 \over 6804 }\, \* \lntwo(4)
          - { 2726824 \over 14175 }\, \* \lntwo(5)
          + { 23358602 \over 567 }\, \* a_4^{}
          + { 21814592 \over 945 }\, \* a_5^{}
\nn \\[1mm] & & \mbox{}
          + { 5343385 \over 162 }\, \* \Lt()
          - { 258056 \over 9 }\, \* \z3 \, \* \Lt()
          + { 12547 \over 9 }\, \* \Lt(2)
          + { 1100 \over 3 }\, \* \Lt(3)
          \bigg]
\nn \\[1mm] & & \mbox{\hspn}
       + \nls \, \*  \bigg[
          - { 48073 \over 108 }
          + { 4091305 \over 1296 }\, \* \z3
          - { 576757 \over 540 }\, \* \zss
          - { 230 \over 3 }\, \* \z5
          - { 685 \over 27 }\, \* \lntwo(2)\, \* \z2
\nn \\[1mm] & & \mbox{}
          + { 685 \over 162 }\, \* \lntwo(4)
          + { 2740 \over 27 }\, \* a_4^{}
          - { 42302 \over 27 }\, \* \Lt()
          + { 28297 \over 36 }\, \* \z3 \, \* \Lt()
          - { 5107 \over 54 }\, \* \Lt(2)
          - { 628 \over 9 }\, \* \Lt(3)
          \bigg]
\nn \\[2mm] & & \mbox{\hspn}
       + \nlt \, \*  \bigg[
          - { 270407 \over 5832 }
          + { 844 \over 27 }\, \* \z3
          + { 1924 \over 81 }\, \* \Lt()
          - { 77 \over 27 }\, \* \Lt(2)
          + { 64 \over 27 }\, \* \Lt(3)
          \bigg]
\; ,
\eea
where $\zeta_n$ denotes the values of the Riemann $\zeta$-function and
$a_n = {\rm Li}_{\:\!n}(\frac{1}{2}) = \sum_{k=1}^{\infty} 
(2^{\:\!k} k^{\:\!n})^{-1}$. 
The corresponding expressions for the \MSb\ and OS masses are given by
\bea
  c_{3,\, \overline{\rm MS } }^{} &\! =\! & c_{3,\,\rm SI}^{} 
       - 152\, \* \Lt()
       - \nl \, \*  
          { 128 \over 3 }\, \* \Lt()
 \:\: , \\[1mm]
  c_{4,\, \overline{\rm MS} }^{} &\! =\! & c_{4,\,\rm SI}^{}
          - { 50186 \over 9 }\, \* \Lt()
          - { 12692 \over 3 }\, \* \Lt(2)
       - \nl \, \*  \bigg[ \,
            { 31282 \over 27 }\, \* \Lt()
          + { 8408 \over 9 }\, \* \Lt(2)
          \bigg]
\nn \\[1mm] & & \mbox{\hspn}
       - \nls \, \*  \bigg[ \,
            { 136 \over 27 }\, \* \Lt()
          - { 640 \over 9 }\, \* \Lt(2)
          \bigg]
\eea
and
\bea
  c_{3,\,\rm OS}^{} &\! =\! & c_{3,\,\rm SI}^{}
       + { 608 \over 3 }
       + \nl \, \*  
          { 512 \over 9 }
 \:\: , \\[2mm]
  c_{4,\,\rm OS}^{} &\! =\! & c_{4,\,\rm SI}^{}
\label{c1OS4}
          + { 297587 \over 27 }
          + 1216\, \* \z2
          - { 304 \over 3 }\, \* \z3
          + { 1216 \over 3 }\, \* \lntwo()\, \* \z2
          + 6688\, \* \Lt()
\nn \\[2mm] & & \mbox{\hspn}
       + \nl \, \*  \bigg[
            { 189238 \over 81 }
          + { 416 \over 3 }\, \* \z2
          - { 256 \over 9 }\, \* \z3
          + { 1024 \over 9 }\, \* \lntwo()\, \* \z2
          + 1472\, \* \Lt()
          \bigg]
\nn \\[2mm] & & \mbox{\hspn}
       - \nls \, \*  \bigg[
            { 4352 \over 81 }
          + { 512 \over 9 }\, \* \z2
          + { 1024 \over 9 }\, \* \Lt()
          \bigg]
\; .
\eea

Our first calculation of the second component of eq.~(\ref{GamHgg}), 
$\mathrm{Im}\, \Pi^{GG}$, to N$^4$LO is addressed below; for a typical Feynman 
diagram see the left part of Figure~ \ref{Fig:hgg}. 
The results are presented and combined with $C_1$ to N$^4$LO results for 
$\Gamma_{H \to\,gg}$ in section~\ref{sec:hgg}.

\vspace*{1mm}
\subsection*{Higgs decay to bottom quarks and the $R$-ratio}
 
\begin{figure}[t]
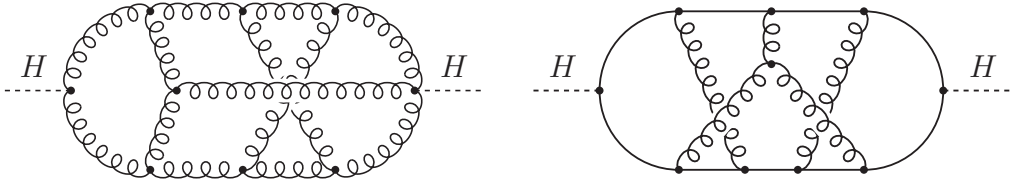

\centering
\begin{axopicture}{(200,80)(0,0)}
\SetWidth{0.6}
\Gluon(135,10)(100,70){3}{8}
\Line[width=9,color=White](135,70)(100,10)
\Gluon(135,70)(100,10){3}{8}
\Line[width=9,color=White](65,40)(165,40)
\GluonArc(65,40)(30,90,180){3}{6}
\GluonArc(65,40)(30,180,270){3}{6}
\GluonArc(135,40)(30,0,90){3}{6}
\GluonArc(135,40)(30,270,360){3}{6}
\Gluon(65,10)(75,40){3}{4}
\Gluon(75,40)(65,70){3}{4}
\Gluon(75,40)(165,40){3}{12}
\Gluon(65,70)(100,70){3}{4}
\Gluon(100,70)(135,70){3}{4}
\Gluon(135,10)(100,10){3}{4}
\Gluon(100,10)(65,10){3}{4}
\Line[dash,dsize=2](10,40)(35,40)
\Line[dash,dsize=2](165,40)(190,40)
\Vertex(35,40){1.5}
\Vertex(165,40){1.5}
\Vertex(65,70){1.5}
\Vertex(100,70){1.5}
\Vertex(135,70){1.5}
\Vertex(65,10){1.5}
\Vertex(100,10){1.5}
\Vertex(135,10){1.5}
\Vertex(75,40){1.5}
\Text(21,45)[b]{$H$}
\Text(180,45)[b]{$H$}
\end{axopicture}
\begin{axopicture}{(200,80)(0,0)}
\SetWidth{0.7}
\Gluon(90,10)(65,70){3}{8}
\Gluon(135,70)(110,10){3}{8}
\Line[width=9,color=White](65,10)(100,50)
\Line[width=9,color=White](100,50)(135,10)
\Arc(65,40)(30,90,180)
\Arc(65,40)(30,180,270)
\Arc(135,40)(30,0,90)
\Arc(135,40)(30,270,360)
\Line(100,70)(65,70)
\Line(135,70)(100,70)
\Line(65,10)(90,10)
\Line(90,10)(110,10)
\Line(110,10)(135,10)
\Gluon(100,50)(100,70){3}{2}
\Gluon(65,10)(100,50){3}{6}
\Gluon(100,50)(135,10){3}{6}
\Line[dash,dsize=2](10,40)(35,40)
\Line[dash,dsize=2](165,40)(190,40)
\Vertex(35,40){1.5}
\Vertex(165,40){1.5}
\Vertex(65,70){1.5}
\Vertex(100,70){1.5}
\Vertex(135,70){1.5}
\Vertex(65,10){1.5}
\Vertex(90,10){1.5}
\Vertex(110,10){1.5}
\Vertex(135,10){1.5}
\Vertex(100,50){1.5}
\Text(21,45)[b]{$H$}
\Text(180,45)[b]{$H$}
\end{axopicture}
 \caption{5-loop Feynman diagrams evaluated for the $H\to gg$ and 
 $H\to \bar{b}^{}b$ decay rates.}
\label{Fig:hgg}
\vspace*{2mm}
\end{figure}

As ref.~\cite{HbbN4LO}, we compute the inclusive Higgs decay to bottom quarks 
at N${}^4$LO in the limit of a small bottom mass, keeping only the 
leading term proportional to the Yukawa coupling.
The corresponding partial decay width can be extracted, again via the optical 
theorem, from the imaginary part of the bottom-Yukawa induced Higgs-boson self 
energy $\Pi^{BB}$,
\beq
\label{GamHbb}
  \Gamma_{H\to\, \bar{b}^{}b} 
  \;=\; \frac{G_{\rm F} \MH m_b^2}{4\sqrt{2}\:\!\pi} \; \tilde R(\MHs)
\quad \mbox{ with } \quad  
  \tilde R(s) \;=\; \frac{\mathrm{Im}\,\Pi^{BB}(-s-i\delta)}{2\,\pi s}
\:\: .
\eeq
A diagram contributing to this process is shown in the right part of 
figure \ref{Fig:hgg}.

\begin{figure}[t]
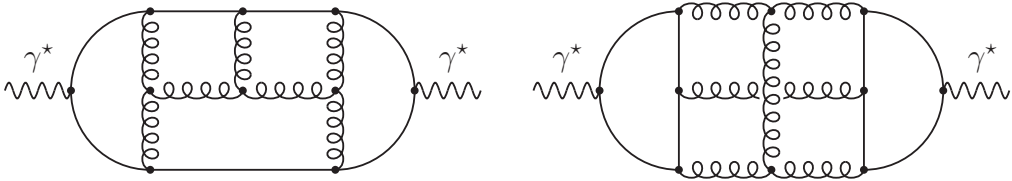

\centering
\begin{axopicture}{(200,80)(0,0)}
\SetWidth{0.7}
\Arc(65,40)(30,90,180)
\Arc(65,40)(30,180,270)
\Arc(135,40)(30,0,90)
\Arc(135,40)(30,270,360)
\Line(100,70)(65,70)
\Line(135,70)(100,70)
\Line(65,10)(135,10)
\Gluon(65,10)(65,40){3}{4}
\Gluon(65,40)(65,70){3}{4}
\Gluon(135,70)(135,40){3}{4}
\Gluon(135,40)(135,10){3}{4}
\Gluon(65,40)(100,40){-3}{4}
\Gluon(100,40)(135,40){-3}{4}
\Gluon(100,40)(100,70){3}{4}
\Photon(10,40)(35,40){3}{4}
\Photon(165,40)(190,40){3}{4}
\Vertex(35,40){1.5}
\Vertex(165,40){1.5}
\Vertex(65,70){1.5}
\Vertex(100,70){1.5}
\Vertex(135,70){1.5}
\Vertex(65,40){1.5}
\Vertex(100,40){1.5}
\Vertex(135,40){1.5}
\Vertex(65,10){1.5}
\Vertex(135,10){1.5}
\Text(23,47)[b]{$\gamma^\star$}
\Text(180,47)[b]{$\gamma^\star$}
\end{axopicture}
\begin{axopicture}{(200,80)(0,0)}
\SetWidth{0.7}
\Gluon(65,40)(135,40){-3}{9}
\Line[width=9,color=White](100,70)(100,10)
\Arc(65,40)(30,90,180)
\Arc(65,40)(30,180,270)
\Arc(135,40)(30,0,90)
\Arc(135,40)(30,270,360)
\Line(65,10)(65,40)
\Line(65,40)(65,70)
\Line(135,70)(135,40)
\Line(135,40)(135,10)
\Gluon(65,70)(100,70){3}{4}
\Gluon(100,70)(135,70){3}{4}
\Gluon(135,10)(100,10){3}{4}
\Gluon(100,10)(65,10){3}{4}
\Gluon(100,10)(100,70){3}{8}
\Photon(10,40)(35,40){3}{4}
\Photon(165,40)(190,40){3}{4}
\Vertex(35,40){1.5}
\Vertex(165,40){1.5}
\Vertex(65,70){1.5}
\Vertex(100,70){1.5}
\Vertex(135,70){1.5}
\Vertex(65,40){1.5}
\Vertex(100,10){1.5}
\Vertex(135,40){1.5}
\Vertex(65,10){1.5}
\Vertex(135,10){1.5}
\Text(23,47)[b]{$\gamma^\star$}
\Text(180,47)[b]{$\gamma^\star$}
\end{axopicture}
\caption{Sample non-singlet (left) and singlet (right) Feynman diagrams
 for which the $1/\ep$ pole terms were computed in our re-calculation of
 the  electromagnetic $R$-ratio at N${}^{4}$LO.}
\label{Fig:Rratio}
\vspace*{2mm}
\end{figure}

The third observable we consider is the hadronic R-ratio, see refs.~\cite
{Rratio1,Rratio2,Rratio3,RratioRC} and references therein, defined as 
\beq
\label{Rratio}
  R(s) \;=\; 
  \frac{\sigma_{e^+e^-\to\, \mathrm{hadrons}}}
       {\sigma_{e^+e^-\to\, \mu^+\mu^-}}\,.
\eeq
Away from the $Z$-pole, the most important contribution to $R(s)$ is given by 
the partial decay width of an off-shell photon into massless quarks. Here we 
re-compute the N${}^4$LO QCD corrections to this electromagnetic contribution. 
Analogous to the Higgs decay, this quantity can be extracted from the 
imaginary part of the photon self energy 
\beq
  \Pi^{\mu\nu}(q^2) \;=\; (-g^{\mu\nu}q^2+q^\mu q^\nu)\,\Pi(q^2)
\eeq
via
\beq
\label{RratEM}
  R^{\,\rm e.m.}(s) \;=\; 12\:\!\pi\, \mathrm{Im}\,\Pi(-s-i\delta) \;=\; N_R 
  \,\bigg[ \Big(\sum_f e_f^2\Big)\;r(s)\, + \Big(\sum_f e_f\Big)^2\: 
  r_{\rm S}^{}(s)  
  \,\bigg]
\eeq
with $N_R = 3$ in QCD. 
The sum runs over $\nf$ quark flavours $f$ with electromagnetic charges 
$e_f$. The functions $r(s)$ and $r_{\rm S}^{}(s)$ represent the respective 
non-singlet and singlet contributions to the $R$-ratio. Example diagrams 
for these two contributions are shown in Figure~\ref{Fig:Rratio}.

\subsection*{Calculations}

For all three observables under consideration, we are interested in 
the imaginary parts of self energies. These can be readily obtained by 
analytic continuation,
\beq
\label{AnCont}
  \mbox{Im } \Pi(-q^2-i\delta) \;=\; \mbox{Im } e^{i \pi \ep L\,} \Pi(q^2) 
  \;=\;  \sin(L\pi \ep)\, \Pi(q^2)
\;,
\eeq
where $\ep=\frac{1}{2}\,(4-D)$ is the dimensional regulator and $L$ the 
number of loops. The crucial point is now that the imaginary part of the 
self energy is suppressed by a factor of $\ep\:\!$:
\beq
  \sin(L\pi \ep) \;=\; L \pi \ep \Big(1 - \frct{1}{3!}\, (L \pi \ep)^2 
  + \frct{1}{5!}\, (L \pi \ep)^4 + \,\ldots \Big) 
\; .
\eeq
Consequently the finite part of $\mbox{Im } \Pi(-q^2)$ can be obtained from 
the $1/\ep$ term of $\Pi(q^2)$.

To compute the single poles we employ the $R^*$-operation for Feynman 
diagrams with arbitrary numerators~\cite{HerzogRuijl17} to express the poles 
of five-loop diagrams in terms of four-loop diagrams.
The $R^*$-operation thus allows us to compute all ingredients required here
using the {\sc Forcer} program \cite{tuACAT2016,tuLL2016,FORCER}, which 
automates the reduction and calculation of massless four-loop self energy 
diagrams.  The same approach was used in~ref.~\cite{beta4b} to compute the 
five-loop beta function for an arbitrary simple compact gauge group.

However, the Higgs decay to gluons poses a much greater computational 
challenge: the diagrams are all quartically divergent. 
In order to infrared rearrange the diagrams, the superficial degree 
of divergence of the diagrams must be logarithmic.
We achieve this by computing the fourth order coefficient of the Taylor 
expansion 
in the external momentum~$q$ about the point $q=0$, i.e. we apply the differential operator
\beq
\frac{1}{4!}q^{\mu_1}q^{\mu_2}q^{\mu_3}q^{\mu_4} \frac{\partial}{\partial q^{\mu_1}}\frac{\partial}{\partial q^{\mu_2}}
\frac{\partial}{\partial q^{\mu_3}}\frac{\partial}{\partial q^{\mu_4}} \left(\quad \bullet \quad \right)\bigg|_{q=0} \,
\eeq
to all Feynman diagrams.
As~a result, an `explosion' of terms, with complicated numerator structures, 
is created. To~deal with this complexity, we have significantly improved our
algorithms, in particular for the reduction of high rank tensor vacuum graphs. 

The Feynman diagrams for all three cases have been generated using QGRAF 
\cite{QGRAF} and were then processed by a {\sc Form} \cite{FORM3,TFORM,FORM4} 
program that assigns the topology and determines the colour factor using the 
program of ref.~\cite{Colour}. 
Diagrams of the same topology, colour factor, and maximal power of $n_\ell$ 
have been combined to meta diagrams for computational efficiency. Lower-order 
self-energy insertions have been treated as described in ref.~\cite{jvLL2016}.

In the case of $\Pi^{\,GG}$, this procedure leads to 1 one-loop, 5 two-loop,
38 three-loop, 394 four-loop and 6405 five-loop meta diagrams. 
These are fewer meta diagrams than for our calculation of the 5-loop beta
function using the background-field method (by a factor $0.68$ to $0.69$ beyond 
two loops), but the present diagrams are much harder, as discussed above.
The computations were performed on the same set of a modern and somewhat 
dated machines as used for our five-loop beta function \cite{beta4b}, 
and required an order of magnitude more time. 
 
In the much more modest cases of $\Pi^{BB}$ and $\Pi$ in eqs.~(\ref{GamHbb})
and (\ref{RratEM}), for which we can use the same diagram set which different
external vertices and projections, we computed 1~one-loop, 2 two-loop, 
9 three-loop, 64 four-loop and 804 five-loop meta diagrams.

We have checked our results by computing all diagrams by at least two different
infrared rearrangements. A different rearrangement results in the computation 
of a different set of counterterms, but should give the same result in the end.
This therefore constitutes a highly non-trivial consistency check of our setup.

The first strategy of IR rearrangement consists of attaching external momenta 
around the line with the worst IR divergence, for example
\begin{equation}
\raisebox{-37pt}{
\begin{axopicture}{(100,80)(-40,-40)}
\Line(-30,0)(-50,0)
\Line(30,0)(50,0) 
\CArc(0,0)(30,0,360)
\Vertex(-30,0){1.5}
\Vertex(30,0){1.5}
\Vertex(0,0){1.5}
\Vertex(0,-30){1.5}
\Line(0,-30)(0,0)
\Line(0,0)(21.21,21.21)
\Line(0,0)(-21.21,21.21)
\CArc(-30,-30)(30,0,90)
\CArc(30,-30)(30,90,180)
\Vertex(21.21,21.21){1.5}
\Vertex(-21.21,21.21){1.5}
\Vertex(7.76,28.97){1.5}
\Vertex(-7.76,28.97){1.5}
\Text(-35,13) {$\mu \nu$}
\Text(-28,-28) {$\nu$}
\Text(28,-28) {$\mu$}
\end{axopicture}
}
\to\quad
\raisebox{-37pt}{
\begin{axopicture}{(100,80)(-40,-40)}
\CArc(0,0)(30,0,360)
\Vertex(-30,0){1.5}
\Vertex(30,0){1.5}
\Vertex(0,0){1.5}
\Vertex(0,-30){1.5}
\Line(0,-30)(0,0)
\Line(0,0)(30,30)
\Line(0,0)(-30,30)
\CArc(-30,-30)(30,0,90)
\CArc(30,-30)(30,90,180)
\Vertex(21.21,21.21){1.5}
\Vertex(-21.21,21.21){1.5}
\Vertex(7.76,28.97){1.5}
\Vertex(-7.76,28.97){1.5}
\Text(-35,13) {$\mu \nu$}
\Text(-28,-28) {$\nu$}
\Text(28,-28) {$\mu$}
\end{axopicture}
}
.
\end{equation}
The resulting integral is an $L$-loop `carpet' integral, which can be reduced 
to a $L-\!1$ loop propagator integral~\cite{FORCER}.
By attaching the external momenta around the worst IR divergent line, the 
number of counterterms that include this line is limited.

The second IR rearrangement consists of inserting a mass into the worst IR 
divergent propagator: 
\begin{equation}
\raisebox{-37pt}{
\begin{axopicture}{(100,80)(-40,-40)}
\Line(-30,0)(-50,0)
\Line(30,0)(50,0) 
\CArc(0,0)(30,0,360)
\Vertex(-30,0){1.5}
\Vertex(30,0){1.5}
\Vertex(0,0){1.5}
\Vertex(0,-30){1.5}
\Line(0,-30)(0,0)
\Line(0,0)(21.21,21.21)
\Line(0,0)(-21.21,21.21)
\CArc(-30,-30)(30,0,90)
\CArc(30,-30)(30,90,180)
\Vertex(21.21,21.21){1.5}
\Vertex(-21.21,21.21){1.5}
\Vertex(7.76,28.97){1.5}
\Vertex(-7.76,28.97){1.5}
\Text(-35,13) {$\mu \nu$}
\Text(-28,-28) {$\nu$}
\Text(28,-28) {$\mu$}
\end{axopicture}
}
\to\quad
\raisebox{-37pt}{
\begin{axopicture}{(100,80)(-40,-40)}
\CArc(0,0)(30,135,405)
\Vertex(-30,0){1.5}
\Vertex(30,0){1.5}
\Vertex(0,0){1.5}
\Vertex(0,-30){1.5}
\Line(0,-30)(0,0)
\Line(0,0)(21.21,21.21)
\Line(0,0)(-21.21,21.21)
\CArc(-30,-30)(30,0,90)
\CArc(30,-30)(30,90,180)
\DoubleArc(0,0)(30,45,135){1}
\Vertex(21.21,21.21){1.5}
\Vertex(-21.21,21.21){1.5}
\Vertex(7.76,28.97){1.5}
\Vertex(-7.76,28.97){1.5}
\Text(-35,13) {$\mu \nu$}
\Text(-28,-28) {$\nu$}
\Text(28,-28) {$\mu$}
\end{axopicture}
}
.
\end{equation}
The resulting counterterm diagrams can always be split up into a massive 
one-loop vacuum bubble and an $L-1$ loop massless propagator integral 
which can be computed using {\sc Forcer}. 
The advantage of this method is that the massive line cannot be part of 
any IR counterterm and that the `carpet' rule reduction is avoided.
Overall, this rearrangement is about 20\% to 50\% 
faster than attaching external momenta.

%
\medskip
\section{Higgs decay to gluons}
\label{sec:hgg}
\setcounter{equation}{0}

After the calculation of the Feynman diagrams, the extraction of the
absorptive part and its renormalization, the coefficients $g_n$ up to N$^4$LO in
\bea
\label{ImGGexp}
  \frac{4\pi}{N_{\!A}\:\!q^4}\, \mbox{Im}\, \Pi^{\,GG}(q^2) \;\equiv\; G(q^2)
  \;=\;  1 + \sum_{n=1} g_n^{} \ar(n) 
\; ,
\eea
with $N_{\!A} = 8$ in QCD, are found to be
\bea
\label{Hgg1}
  g_1^{} &\! =\! &
     {73 \over 3}\, \* \ca
   - {14 \over 3}\, \* \nl
 \:\: , \\[2mm]
\label{Hgg2}
  g_2^{} &\! =\! &
       \cas \: \* \Bigg[
          {37631 \over 54}
        - {242 \over 3}\, \* \z2
        - 110\, \* \z3
        \Bigg]
     \: - \: \ca \* \,\nl \: \* \Bigg[
          {6665 \over 27}
        - {88 \over 3}\, \* \z2
        + 4\, \* \z3
        \Bigg]
\nn \\ & & \mbox{}
     - \: \cf \* \,\nl \, \* \Bigg[
          {131 \over 3}
        - 24\, \* \z3
        \Bigg]
     \: + \: \nls \: \* \Bigg[
          {508 \over 27}
        - {8 \over 3}\, \* \z2
        \Bigg]
 \:\: ,\\[2mm]
\label{Hgg3} 
  g_3^{} &\! =\! &
         \cat \, \* \Bigg[ \,
            {15420961 \over 729}
          - {45056 \over 9}\,\* \z2
          - {178156 \over 27}\,\* \z3
          + {3080 \over 3}\, \* \z5
          \Bigg]
\nn \\ & & \mbox{}
       - \: \cas \* \,\nl \, \* \Bigg[ \,
            {2670508 \over 243}
          - {8084 \over 3}\, \* \z2
          - {9772 \over 9}\, \* \z3
          + {80 \over 3}\, \* \z5
          \Bigg]
\nn \\ & & \mbox{}
       - \: \cf \* \ca \* \,\nl\, \* \Bigg[ \,
            {23221 \over 9}
          - {572 \over 3}\, \* \z2
          - 1364\, \* \z3
          - 160\, \* \z5
          \Bigg]
\nn \\ & & \mbox{}
       + \: \cfs \* \,\nl \, \* \Bigg[ \,
            {221 \over 3}
          + 192\, \* \z3
          - 320\, \* \z5
          \Bigg]
       \: + \: \ca \* \,\nls \, \* \Bigg[ \,
            {413308 \over 243}
          - {1384 \over 3}\, \* \z2
          + {56 \over 9}\, \* \z3
          \Bigg]
\qquad \nn \\ & & \mbox{}
       + \: \cf \* \,\nls \, \* \Big[ 
            440\:
          - {104 \over 3}\, \* \z2
          - 240\, \* \z3
          \Big]
       \: - \: \nlt \, \* \Bigg[ \,
            {57016 \over 729}
          - {224 \over 9}\, \* \z2
          - {64 \over 27}\, \* \z3
          \Bigg]
\qquad
\eea
and
\bea
\label{Hgg4}
  \hspace*{2mm} g_4^{} &\!\! =\!\! &
         \caf \, \* \Bigg[ \,
            { 5974862279 \over 8748 }
          - { 58922654 \over 243 }\, \* \z2
          - { 25166402 \over 81 }\, \* \z3
          + { 292556 \over 45 }\, \* \zss
          + { 266200 \over 9 }\, \* \z2 \* \z3
\nn \\ & & \mbox{\hspp}
          + { 1817200 \over 27 }\, \* \z5
          + { 121000 \over 9 }\, \* \zts
          - { 96250 \over 9 }\, \* \z7
          \Bigg]
\nn \\ & & \mbox{\hspn\hspn}
       - \, \dfAAna \, \*  \Bigg[ \,
            { 6416 \over 27 }
          - { 54160 \over 9 }\, \* \z3
          - { 1408 \over 5 }\, \* \zss
          + { 13760 \over 3 }\, \* \z5
          - { 19360 \over 3 }\, \* \zts
          + { 6160 \over 3 }\, \* \z7
          \Bigg]
\nn \\ & & \mbox{\hspn\hspn}
       - \, \cat\, \* \nl \, \*  \Bigg[ \,
           { 1025827736 \over 2187 }
         - { 41587004 \over 243 }\, \* \z2
         - { 8812352 \over 81 }\, \* \z3
         + { 211736 \over 45 }\, \* \zss
         + 9680\, \* \z2 \* \z3
\nn \\ & & \mbox{\hspp}
         + { 109220 \over 9 }\, \* \z5
         - { 8800 \over 9 }\, \* \zts
         + { 3500 \over 9 }\, \* \z7
          \Bigg]
\nn \\ & & \mbox{\hspn\hspn}
       - \, \cas \* \cf\, \* \nl \, \*  \Bigg[ \,
           { 348948545 \over 2916 }
         - 22340\, \* \z2
         - { 1869710 \over 27 }\, \* \z3
         + { 656 \over 15 }\, \* \zss
         + { 19360 \over 3 }\, \* \z2 \* \z3
\nn \\ & & \mbox{\hspp}
         - { 35540 \over 3 }\, \* \z5
         + { 17600 \over 3 }\, \* \zts
         - { 7000 \over 3 }\, \* \z7
          \Bigg]
\nn \\ & & \mbox{\hspn\hspn}
       + \, \ca \* \cfs\, \* \nl \, \*  \Bigg[ \,
            { 609521 \over 162 }
          - { 484 \over 3 }\, \* \z2
          + { 450374 \over 27 }\, \* \z3
          + { 352 \over 15 }\, \* \zss
          - { 63040 \over 3 }\, \* \z5
          - 5600\, \* \z7
          \Bigg]
\nn \\ & & \mbox{\hspn\hspn}
       + \, \cft\, \* \nl \, \*  \Bigg[ \,
            { 1034 \over 3 }
          - 388\, \* \z3
          - 4560\, \* \z5
          + 5600\, \* \z7
          \Bigg]
\nn \\ & & \mbox{\hspn\hspn}
       + \, \dfFAna\, \* \nl \, \*  \Bigg[ \,
            { 44864 \over 27 }
          - { 140128 \over 9 }\, \* \z3
          - { 3328 \over 5 }\, \* \zss
          + { 20800 \over 3 }\, \* \z5
          - { 14080 \over 3 }\, \* \zts
          + { 2240 \over 3 }\, \* \z7
          \Bigg]
\nn \\ & & \mbox{\hspn\hspn}
       + \, \cas\, \* \nls \*  \Bigg[ \,
            { 26855351 \over 243 }
          - { 3479386 \over 81 }\, \* \z2
          - { 83536 \over 9 }\, \* \z3
          + { 19472 \over 15 }\, \* \zss
          + { 1760 \over 3 }\, \* \z2 \* \z3
          - { 1240 \over 9 }\, \* \z5
          + { 160 \over 9 }\, \* \zts
          \Bigg]
\nn \\ & & \mbox{\hspn\hspn}
       + \, \cf \* \ca\, \* \nls \, \*  \Bigg[ \,
            { 29816212 \over 729 }
          - { 71888 \over 9 }\, \* \z2
          - { 563948 \over 27 }\, \* \z3
          + { 224 \over 15 }\, \* \zss
          + { 7040 \over 3 }\, \* \z2 \* \z3
          - { 7000 \over 3 }\, \* \z5
          - { 640 \over 3 }\, \* \zts
          \Bigg]
\nn \\ & & \mbox{\hspn\hspn}
       + \, \cfs\, \* \nls \, \*  \Bigg[ \,
            { 90491 \over 81 }
          - { 200 \over 3 }\, \* \z2
          - { 138968 \over 27 }\, \* \z3
          - { 352 \over 15 }\, \* \zss
          + 4400\, \* \z5
          + 640\, \* \zts
          \Bigg]
\nn \\ & & \mbox{\hspn\hspn}
       - \, \dfFFna\, \* \nls \, \*  \Bigg[ \,
           { 68096 \over 27 }
         - { 39424 \over 9 }\, \* \z3
         - { 1024 \over 5 }\, \* \zss
         + 1280\, \* \z5
         - { 2560 \over 3 }\, \* \zts
          \Bigg]
\nn \\ & & \mbox{\hspn\hspn}
       - \, \ca\, \* \nlt \, \*  \Bigg[ \,
           { 46491973 \over 4374 }
         - { 1099028 \over 243 }\, \* \z2
         - { 23720 \over 81 }\, \* \z3
         + { 1408 \over 9 }\, \* \zss
         - { 320 \over 9 }\, \* \z2 \* \z3
         - { 800 \over 27 }\, \* \z5
          \Bigg]
\nn \\ & & \mbox{\hspn\hspn}
       - \, \cf\, \* \nlt \, \*  \Bigg[ \,
           { 2282351 \over 729 }
         - { 6224 \over 9 }\, \* \z2
         - { 5200 \over 3 }\, \* \z3
         + { 640 \over 3 }\, \* \z2 \* \z3
          \Bigg]
\nn \\ & & \mbox{\hspn\hspn}
       + \, \nlf \, \*  \Bigg[ \,
            { 773024 \over 2187 }
          - { 40640 \over 243 }\, \* \z2
          - { 2240 \over 81 }\, \* \z3
          + { 64 \over 9 }\, \* \zss
          \Bigg]
\:\: .
\eea
Eqs.~(\ref{Hgg1}) -- (\ref{Hgg3}) agree with the previous results in 
refs.~\cite{Inami:1982xt,Djouadi:1991tka,HggNNLO,HggN3LO,MV2,Davies:2017rle} 
($n_\ell$ instead of $\nf$ is often used for the number of light flavours)
eq.~(\ref{Hgg4}) represents the main new result of the present article.
In all these equations $\tf = 1/2$ has been inserted; this factor can be
re-instated by substituting $\nl \ra 2\,\tf\:\!\nl$ in all terms that do
not involve quartic group invariants. 

The coefficients (\ref{Hgg1}) -- (\ref{Hgg4}) are valid for the standard 
choice $\mu^2 = q^2$ of the renormalization scale. 
The additional terms for $\mu^2 \neq q^2$ can be obtained from the scale 
invariance of $(\beta(\ars)/\ars)^2 \, \mbox{Im}\, \Pi^{\,GG}(q^2)$
\cite{Inami:1982xt,HggN3LO}. 
This can be done, e.g., by inserting the expansion of $\ars(q^2)$ in 
terms of $\ars(\mu^2)$ which can be read off to the order required here, 
for example, from eq.~(2.9) and footnote 2 of ref.~\cite{NV3}. 
The resulting generalizations of eqs.~(\ref{Hgg1}) -- (\ref{Hgg4}) read 
\bea
\label{Hgg1mu}
  \g1(L_q) &\! =\! & 
       g_1^{} 
       \,-\, 2\*\b0 \*\, L_q  
 \:\: , \nn \\[0.5mm]
\label{Hgg2mu}
  \g2(L_q) &\! =\! &
       g_2{} 
       \,-\, ( 4\*\b1 + 3\*\b0\*\g1 ) \* L_q 
       \,+\, 3\*\B(0,2) \* \, L_q^2
 \:\: , \nn \\[1mm]
\label{Hgg3mu}
  \g3(L_q) &\! =\! & 
       \g3 
       \,-\, ( 6\*\b2 + 5\*\b1\*\g1 + 4\*\b0\*\g2 ) \* L_q
       \,+\, ( 13\*\b0\*\b1 + 6\*\B(0,2)\*\g1 ) \* L_q^2
       \,-\, 4\*\*\B(0,3) \*\, L_q^3 
 \:\: , \nn \\[1mm]
\label{Hgg4mu}
  \g4(L_q) &\! =\! & 
       \g4 
       \,-\, ( 8\*\b3 + 7\*\b2\*\g1 + 6\*\b1\*\g2 + 5\*\b0\*\g3 ) \* L_q
\nn \\[0.5mm] & & \quad\mbox{}
       \,+\, ( 12\*\B(1,2) + 22\*\b0\*\b2 
             + 43/2\*\b0\*\b1\*\g1 + 10\*\B(0,2)\*\g2 ) \* L_q^2
\nn \\[0.5mm] & & \quad\mbox{}
       \,-\, ( 83/3\*\B(0,2)\*\b1 + 10\*\B(0,3)\*\g1 ) \* L_q^3
       \,+\, 5\*\B(0,4) \*\, L_q^4
\eea
in terms of the above coefficients $g_{n}^{}$, the coefficients $\b{n}$ 
of the \MSb~beta function up to N$^3$LO \cite{beta3a,beta3b}
and $L_q\,\equiv\,\ln (q^2/\mu^2)$.
The resulting explicit coefficients up to $g_{3}^{}$ agree with eq.~(26) 
of ref.~\cite{DaviesSW17}, where the definitions of $L$ and $\ars$ are
slightly different.

At the scale $\mu^2 = q^2$ the numerical expansion of the function $G(q^2)$ 
in  eq.~(\ref{ImGGexp}) is given~by 
\bea
\label{HggNumNf}
  \nl = 1 &\!\! : &
       1 + 5.4377939\,\als   + 20.720313\,\as(2)
         + 58.92184 \,\as(3) + 118.0078\,\as(4) + \ldots 
\; , \nn \\[0.5mm]
  \nl = 3 &\!\! : &
       1 + 4.6950708\,\als   + 13.472440\,\as(2)
         + 20.66395 \,\as(3) - 15.96239 \,\as(4) + \ldots
\; , \nn \\[0.5mm]
  \nl = 5 &\!\! : &
       1 + 3.9523478\,\als   + 6.9555141\,\as(2)
         - 6.851753 \,\as(3) - 75.25914 \,\as(4) + \ldots
\; , \nn \\[0.5mm]
  \nl = 7 &\!\! : &
       1 + 3.2096247\,\als   + 1.1695355\,\as(2)
         - 24.45788 \,\as(3) - 76.99773 \,\as(4) + \ldots
\; , \nn \\[0.5mm]
  \nl = 9 &\!\! : &
       1 + 2.4669016\,\als   - 3.8854956\,\as(2)
         - 32.98703 \,\as(3) - 37.30247 \,\as(4) + \ldots
\;\quad
\eea
for QCD with up to 5 quark families, i.e., $\nl = 1,\,\,\ldots,\,9$
light flavours. In the only physically relevant case of $\nl = 5$ the
effect of the fourth-order correction is larger than that of the previous
order for $\als \gsim 0.1$. 
It is clear from eqs.~(\ref{HggNumNf}), though, that this is not a 
generic feature of the QCD perturbation series, but a consequence of the 
`accidentally' small size of the third-order term for this number of 
flavours. A similar situation has been observed for Higgs decay to bottom 
quarks, see eq.~(8) of ref.~\cite{HbbN4LO} and eq.~(\ref{HbbNumNf}) below.

The fourth-order coefficient $g_4^{}$ in eq.~(\ref{Hgg4}) is the first to 
receive contributions from quartic group invariants. The overall effect
of these terms is small in the range of $\nl$ considered above; nullifying
all these terms changes the coefficients of $\as(4)$ in eqs.~(\ref{HggNumNf}) 
by about 5\% or less for $\nl \neq 3$. For $\nl=3$, the relative effect is 
larger since the coefficient is atypically small.

As discussed in ref.~\cite{HggN3LO}, the $\z2 = \pi^2/6$ contributions 
in eqs.~(\ref{Hgg2}) -- (\ref{Hgg3}) only arise from the analytic 
continuation (\ref{AnCont}) and are predictable from lower-order results. 
The same holds for the terms linear in $\z2$ in eq.~(\ref{Hgg4}). 
However, the `genuine' five-loop contributions from the functions 
$\Pi^{\,GG}(q^2)$ include terms with $\zss$, so not all powers of
$\pi^2$ are `kinematical' from this order onwards.
The numerical decomposition of the expansion in eq.~(\ref{ImGGexp}) into 
the `genuine' and `kinematical' contributions (underlined below) reads
\bea
\label{ImGGexpZ2}
  G(q^2) &\!=\!&
  1 +   3.952348\,\als  
    + ( 10.629125 - \underline{3.673611}^{}  ) \,\as(2)
\nn \\ & & \;\;\mbox{}
    + ( 28.57606  - \underline{35.42782}^{}  ) \,\as(3)
    + ( 89.55798  - \underline{164.81711}^{} ) \,\as(4)
\quad
\eea
for the physical case of $\nl = 5$. The numbers up to order $\as(3)$ agree,
of course, with ref.~\cite{HggN3LO}; the $\nl$-dependent decomposition of 
our new coefficient $g_4^{}$ in eq.~(\ref{Hgg4}) is 
\bea
  g_4^{} &\!=\!&
      1267.05129 - \underline{1048.43622}
  - ( 394.681626 - \underline{281.704409}^{} ) \,\nl 
\nn \\ & & \mbox{}
  + ( 37.9589880 - \underline{25.1937144}^{} ) \,\nls
  - ( 1.28868582 - \underline{0.89082162}^{} ) \,\nlt
\nn \\ & & \mbox{}
  + ( 0.01284135 - \underline{0.01026045}^{} ) \,\nlf
\:\: .
\eea
The cancellations between the genuine and kinematic contributions are 
somewhat less striking than for the corresponding contribution to $H 
\rightarrow \bar{b}^{}b$, see eq.~(7) of ref.~\cite{HbbN4LO}, yet the 
conclusion remains the same: it is not possible to obtain reliable 
results without computing the genuine contributions.

The decay rate $\Gamma_{H \ra\, gg}$ in the limit of a heavy top 
quark and $\nl$ effectively massless flavours is obtained by combining 
eqs.~(\ref{ImGGexp}) -- (\ref{Hgg4mu}) with the corresponding expansion 
of the coefficient function $C_1$ in eqs.~(\ref{C1exp}) -- (\ref{c1OS4})
above.  The resulting $K$-factors, defined by $\Gamma \,=\, K\:\! 
\Gamma_{\mathrm{Born}}$ at $\mu^2=\MHs$, see eq.~(\ref{GamHggE}) below, 
are given by
\bea
\label{KSInumNf}
  K_{\rm SI^{}}(\nl\!=\!1) &\! =\! &
       1 + 7.188498\,\als    + 32.65167\,\as(2)
         + 112.015 \,\as(3)  + 298.873 \,\as(4) + \ldots
\; , \nn \\[0.5mm]
  K_{\rm SI^{}}(\nl\!=\!3) &\! =\! &
       1 + 6.445775\,\als    + 23.74728\,\as(2)
         + 56.0755 \,\as(3)  + 62.4363 \,\as(4) + \ldots
\; , \nn \\[0.5mm]
  K_{\rm SI^{}}(\nl\!=\!5) &\! =\! &
       1 + 5.703052\,\als    + 15.57384\,\as(2)
         + 12.5520 \,\as(3)  - 72.0916 \,\as(4) + \ldots
\; , \nn \\[0.5mm]
  K_{\rm SI^{}}(\nl\!=\!7) &\! =\! &
       1 + 4.960329\,\als    + 8.131350\,\as(2)
         - 19.3879 \,\as(3)  - 123.853 \,\as(4) + \ldots
\; , \nn \\[0.5mm]
  K_{\rm SI^{}}(\nl\!=\!9) &\! =\! &
       1 + 4.217606\,\als    + 1.419805\,\as(2)
         - 40.5769 \,\as(3)  - 110.998 \,\as(4) + \ldots
\;\quad
\eea
for a scale-invariant top mass $\mu_t = 164$ GeV, and by
\bea
\label{KOSnumNf}
  K_{\rm OS^{}}(\nl\!=\!1) &\! =\! &
       1 + 7.188498\,\als    + 32.61874\,\as(2)
         + 112.031 \,\as(3)  + 300.278 \,\as(4) + \ldots
\; , \nn \\[0.5mm]
  K_{\rm OS^{}}(\nl\!=\!3) &\! =\! &
       1 + 6.445775\,\als    + 23.69992\,\as(2)
         + 56.1329 \,\as(3)  + 64.5259 \,\as(4) + \ldots
\; , \nn \\[0.5mm]
  K_{\rm OS^{}}(\nl\!=\!5) &\! =\! &
       1 + 5.703052\,\als    + 15.51204\,\as(2)
         + 12.6660 \,\as(3)  - 69.3287 \,\as(4) + \ldots
\; , \nn \\[0.5mm]
  K_{\rm OS^{}}(\nl\!=\!7) &\! =\! &
       1 + 4.960329\,\als    + 8.055116\,\as(2)
         - 19.2021 \,\as(3)  - 120.458 \,\as(4) + \ldots
\; , \nn \\[0.5mm]
  K_{\rm OS^{}}(\nl\!=\!9) &\! =\! &
       1 + 4.217606\,\als    + 1.329135\,\as(2)
         - 40.3039 \,\as(3)  - 107.042 \,\as(4) + \ldots
\;\quad
\eea
for an on-shell top mass of $M_t = 173$ GeV. The effect of the coefficient
functions is positive, except for their N$^3$LO and N$^4$LO contributions
at large $\nl$.

The mass- and scale-dependent expansion coefficients $\gamma_n^{}$ for 
the physical case $\nl=5$ in
\beq
\label{GamHggE}
  \Gamma_{H \ra gg} \;=\; \frac{G_{F\,} \MHt}{36\:\!\pi^{\,3}\sqrt{2}}
  \;\sum_{n=0} \gamma_n^{}(\MH, m_t, \mu)  \left( \als(\mu^2) \right)^{n+2}
\eeq
(the $n=0$ contribution is the Born result) are given by $\gamma_0^{} = 1$ and
\bea
  \gamma_{1,\,\rm SI}^{} &\!=\!&
    5.703052 - 1.220188\, L_H
\; , \nn \\[0.5mm] 
  \gamma_{2,\,\rm SI}^{} &\!=\!& 
    15.887961 - 0.578375\, L_{tH}
    - 10.927911\, L_H + 1.116644\, L_H^2
\; , \nn \\[0.5mm]
  \gamma_{3,\,\rm SI}^{} &\!=\!&
    14.59257 - 3.94891\, L_{tH} + 0.352863\, L_{tH}^2
\nn \\ & & \mbox{}
    - ( 43.14427 - 1.41145\, L_{tH} ) L_H
    + 13.78227\, L_H^2 - 0.908344\, L_H^3
\; , \nn \\[0.5mm]
  \gamma_{4,\,\rm SI}^{} &\!=\!& \mbox{}
    - 66.75046 - 11.35498\, L_{tH} 
    +  2.91649\, L_{tH}^2 - 0.215280\, L_{tH}^3
\nn \\ & & \mbox{}
    - ( 62.02230 - 12.61251\, L_{tH} + 1.07640\, L_{tH}^2 ) L_H
\nn \\ & & \mbox{}
    + ( 71.32360 -  2.15280\, L_{tH} ) L_H^2
    - 14.37869\, L_H^3 + 0.692719\, L_H^4
\eea 
with $L_H = \ln (\MHs/\mu^2)$ and $L_{tH} = \ln (\mu_t^2 / \MHs)$
in terms of the scale-invariant top-quark mass. The corresponding OS-mass
coefficient have the same form at NLO and NNLO, and read
\bea
  \gamma_{3,\rm OS}^{} &\!=\!&  \gamma_{3,\,\rm SI}^{} +
  0.490940
\; , \nn \\
\gamma_{4,\rm OS}^{} &\!=\!&  \gamma_{4,\,\rm SI}^{} + 
  4.21311 - 0.89856\, L_{tH} - 1.49760\, L_H
\; ,
\eea
where the top-mass logarithms are now given by $L_{tH} = \ln (M_t^2 /\MHs)$.

\begin{figure}[p]
\vspace*{-3mm}
\centerline{\epsfig{file=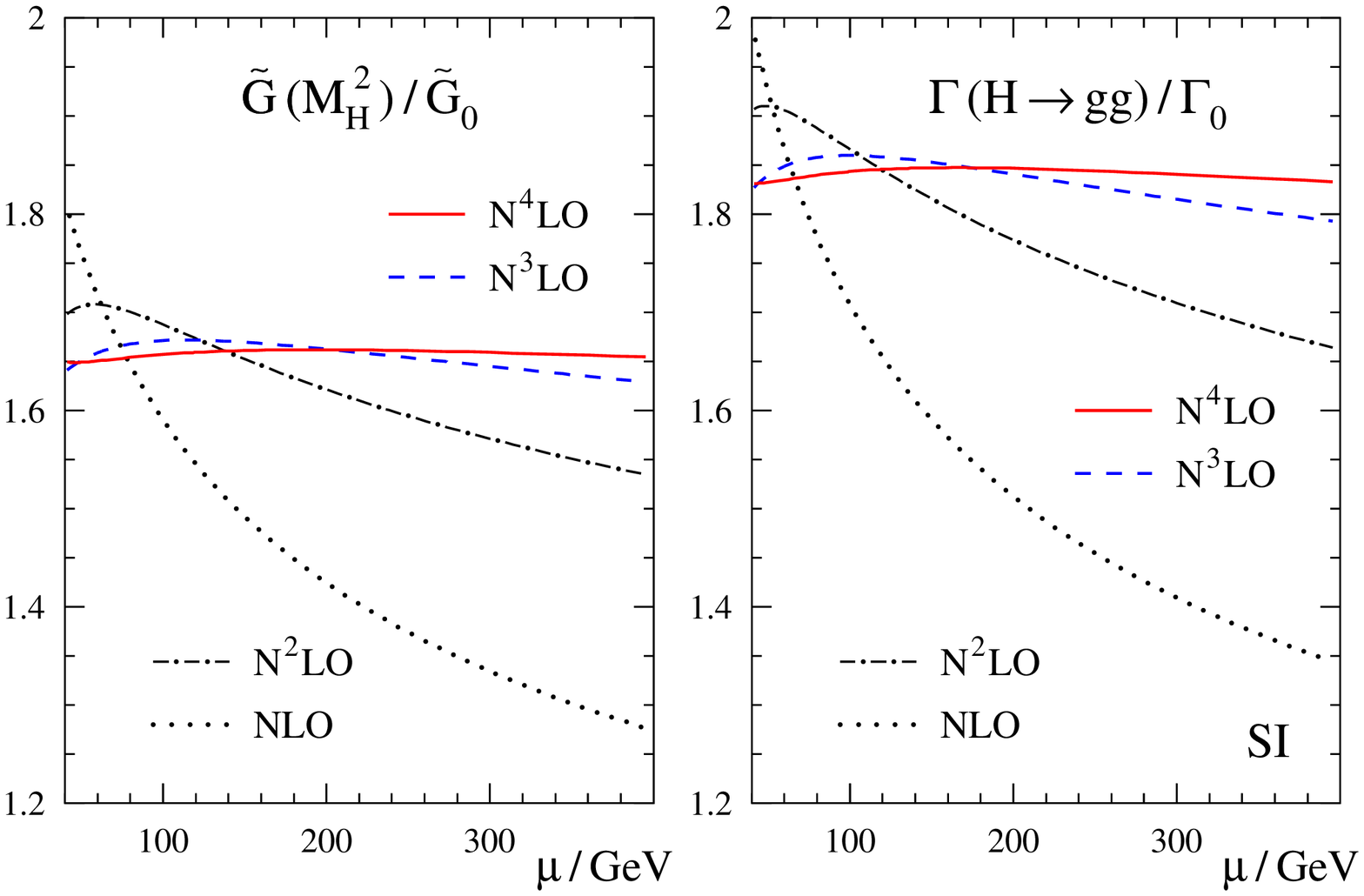,width=15.5cm,angle=0}}
\vspace{-2mm}
\caption{ \label{fig:hggscale} \small 
 The renormalization-scale dependence of $\widetilde{G} = (\beta(\ars)/\ars)^2 
 G(\MHs)$, with $G(q^2)$ defined in eq.~(\ref{ImGGexp}), at $\nl=5$
 (left panel), and of the decay width $\Gamma_{H \ra\, gg}$ (right panel), 
 both normalized as discussed in the text, up to N$^4$LO in \MSb\ for 
 $\als(\MZs) = 0.118$, $\MH = 125 \mbox{ GeV}$ and $\mu_t = 164 \mbox{ GeV}$.}
\vspace{-2mm}
\end{figure}

The size of the higher order corrections and the improvement of the
renormalization scale dependence from NLO to N$^4$LO is illustrated in
Fig.~\ref{fig:hggscale} for $(\beta(\ars)/\ars)^2 \, \mbox{Im}\, 
\Pi^{\,GG}(\MHs)$, recall the discussion above eq.~(\ref{Hgg1mu}), and for
$\Gamma_{H\ra\, gg}$ in eq.~(\ref{GamHggE}). The first term in the expansion
has been normalized for both quantities, i.e., $\Gamma_0$ in the figure is
given by
\beq
  \Gamma_0 = G_F \MHt / (36 \pi^3 \sqrt{2}) \cdot (\als(\MHs))^2
  \quad \mbox{with} \quad \als(\MHs) = 0.11264
\eeq
which corresponds to $\als(\MZs) = 0.118$.
The normalized decay rate is shown for an SI mass of 164 GeV. The very
similar results for an OS mass of 173 GeV are shown below.

\begin{figure}[p]
\vspace*{-1mm}
\centerline{\epsfig{file=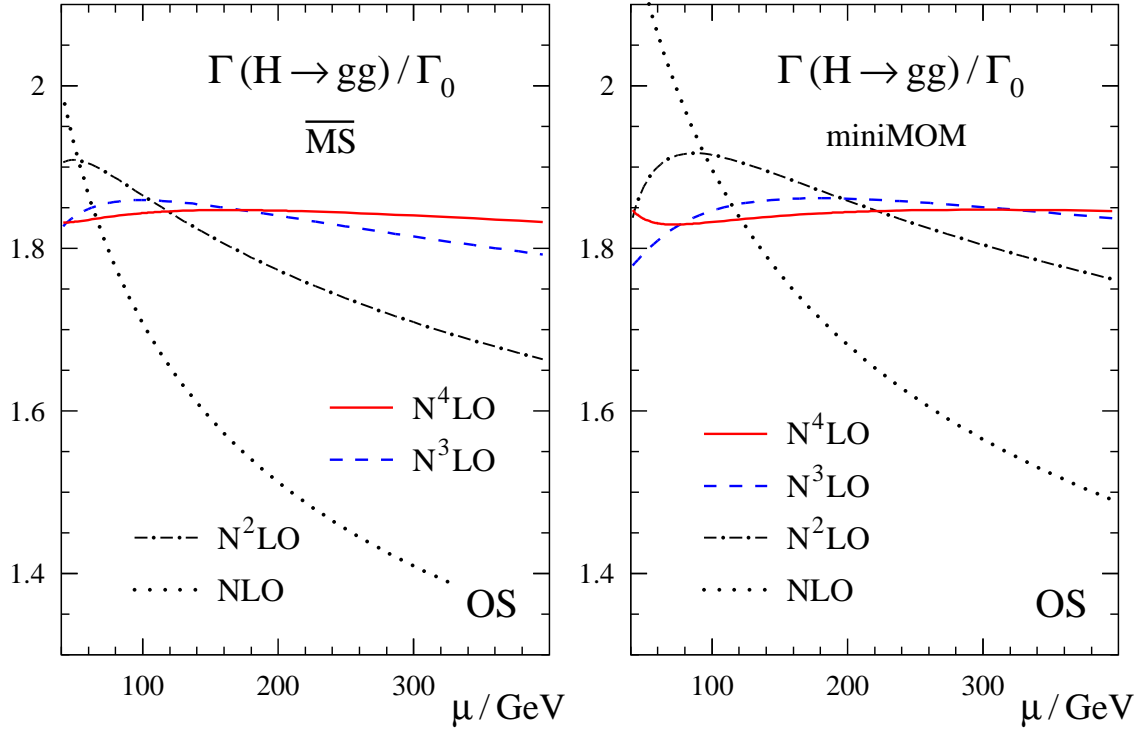,width=15.5cm,angle=0}}
\vspace*{-2mm}
\caption{ \label{fig:hggscale2} \small
 The renormalization scale dependence of the decay width 
 $\Gamma_{H \ra\,gg}$, normalized as the right part of 
 fig.~\ref{fig:hggscale}, for an on-shell top mass of 173 GeV in \MSb\ 
 and the miniMOM scheme.}
\vspace*{-2mm}
\end{figure}

The effect of the N$^4$LO correction to $\Gamma_{H\ra\, gg}$ is $-0.6\%$ at
$\mu = \MH$, and $-0.8\% \:/\, +0.9\%$ at $\mu = 0.5 \:/\, 2\, \MH$, 
respectively. The total N$^4$LO result at $\mu = \MH$ is 1.846$\,\Gamma_0$,
and its range in the above scale interval is (1.836 $-$ 1.848)$\,\Gamma_0$.
The N$^4$LO scale variation between $\mu = 1/3\,\MH$ and $\mu = 3\,\MH$ 
is as small as 0.9\% (full width), a reduction of almost a factor of four 
with respect to the corresponding N$^3$LO result. 
The dependence of $\Gamma_{H\ra\, gg}$ on the top mass is very weak: 
changing $\mu_t$ by 4 GeV \cite{ABMP17} changes the result by only 0.04\%. 
The largest uncertainty at N$^4$LO is due to $\als$: changing $\als(\MZs)$ 
by 1\% changes the result by 2.5\%.

Averaging the result at $\mu = \MH$ and the central value of the above 
scale interval, and using the shift at $\mu = \MH$ from N$^3$LO to N$^4$LO 
(or twice the width of the above scale range) for a conservative estimate 
of the series expansion uncertainty, the N$^4$LO result 
-- without $1/m_t$ corrections and light-quark mass effects -- 
can be summarized as
\bea
\label{GHggRes}
  \Gamma_{\rm N^4LO} (H\ra gg) \;=\; \Gamma_0 \left( 
  1.844 \:\pm\: 0.011_{\,\rm series} \:\pm\: 0.045_{\,\als(\MZ), 1\%}
  \right)
\; .
\eea
The uncertainty due to the truncation of the perturbation series at N$^4$LO is 
definitely much smaller than the uncertainty due to that of $\als(\MZ)$ which 
may exceed the value of 1\% quoted by the Particle Data Group \cite{PDG17}; 
see ref.~\cite{ABMP17} for a recent deviating analysis. 

We conclude our discussion of $\Gamma_{H\ra\,gg}$ by re-expressing its
perturbative expansion in another renormalization scheme, the miniMOM
scheme \cite{MiniMOM1,MiniMOM2}. The transformation from \MSb\ to miniMOM 
and the beta function in this scheme have been derived at N$^4$LO in 
ref.~\cite{N4LOmMOM}. 

The decay width (\ref{GamHggE}) in the OS scheme for the top mass can
be readily transformed by expressing $\als$ in terms of $\alpha_{\rm s,MM}$. 
For the Landau-gauge miniMOM scheme one finds
\bea
\label{KOSnNfMM}
  K_{\rm OS}^{\rm MM}(\nl\!=\!1) &\! =\! &
       1 + 5.123905\,\als    + 10.56499\,\as(2)
         - 7.47722 \,\as(3)  - 112.155 \,\as(4) + \ldots
\; , \nn \\[1mm]
  K_{\rm OS}^{\rm MM}(\nl\!=\!3) &\! =\! &
       1 + 4.734860\,\als    + 7.406951\,\as(2)
         - 14.9763 \,\as(3)  - 91.2437 \,\as(4) + \ldots
\; , \nn \\[1mm]
  K_{\rm OS}^{\rm MM}(\nl\!=\!5) &\! =\! &
       1 + 4.345814\,\als    + 4.379443\,\as(2)
         - 21.5506 \,\as(3)  - 71.9231 \,\as(4) + \ldots
\; , \nn \\[1mm]
  K_{\rm OS}^{\rm MM}(\nl\!=\!7) &\! =\! &
       1 + 3.956769\,\als    + 1.482460\,\as(2)
         - 27.1850 \,\as(3)  - 53.7325 \,\as(4) + \ldots
\; , \nn \\[1mm]
  K_{\rm OS}^{\rm MM}(\nl\!=\!9) &\! =\! &
       1 + 3.567723\,\als    - 1.283997\,\as(2)
         - 31.8645 \,\as(3)  - 36.2907 \,\as(4) + \ldots
\;\quad
\eea
in terms of $\als = \alpha_{\rm s,\,MM}$ (here)
at $\mu = \MH$ for $M_t = 173 \mbox{ GeV}$. The miniMOM version of the
$\nl=5$ OS-scheme expansion coefficients in eq.~(\ref{GamHggE}) is given by
\bea
\label{gamOSMM}
  \gamma_{1,\,\rm OS}^{\,\rm MM} &\!=\!&
    4.345814 - 1.220188\, L_H
\; , \nn \\[1mm]
  \gamma_{2,\,\rm OS}^{\,\rm MM} &\!=\!&
    4.755361 - 0.578375\, L_{tH}
    - 8.443784\, L_H + 1.116644\, L_H^2
\; , \nn \\[1mm]
  \gamma_{3,\,\rm OS}^{\,\rm MM} &\!=\!&
    - 20.15349 - 2.37892\, L_{tH} + 0.352863\, L_{tH}^2
\nn \\ & & \mbox{}
    - ( 15.98471 - 1.41145\, L_{tH} ) L_H
    + 10.75116\, L_H^2 - 0.908344\, L_H^3
\; , \nn \\[1mm]
  \gamma_{4,\,\rm OS}^{\,\rm MM} &\!=\!& \mbox{}
    - 72.28293 - 0.47286\, L_{tH}
    +  1.71919\, L_{tH}^2 - 0.215280\, L_{tH}^3
\nn \\ & & \mbox{}
    + ( 52.95134 + 7.82332\, L_{tH} + 1.07640\, L_{tH}^2 ) L_H
\nn \\ & & \mbox{}
    + ( 27.68353 -  2.15280\, L_{tH} ) L_H^2
    - 11.29660\, L_H^3 + 0.692719\, L_H^4
\eea
The value of the strong coupling in this miniMOM scheme is 
larger than that in \MSb\ with $\alsMM(\MZs) = 1.0960\,\als(\MZs)$ for 
$\als(\MZs) = 0.118$ \cite{N4LOmMOM}, or more generally for $\nl = 5$:
\bea
  \alsMM &\!=\!& \als
   + 0.67862\,\as(2)
   + 0.91231\,\as(3)
   + 1.5961 \,\as(4)
   + 3.1629 \,\as(5)
   + {\cal O}(\as(6)) 
\; .
\eea
This is compensated by lower-order coefficients in eqs.~(\ref{KOSnNfMM}) and
(\ref{gamOSMM}) that are smaller than their \MSb\ counterparts. The N$^3$LO 
and N$^4$LO terms for $\nl=5$ are not smaller, though. 
  
The resulting perturbative expansion of $\Gamma_{H \ra\, gg}$ in the
Landau-gauge miniMOM scheme is shown in fig.~\ref{fig:hggscale2}. The
general pattern in miniMOM is somewhat different from that in \MSb\
-- qualitatively the curves appear shifted to the right.
Yet the overall scale range for the interval in $\mu$ displayed in the figure
is very similar to (if slightly wider than) that in the \MSb\ scheme and
covered by eq.~(\ref{GHggRes}).
Given this small uncertainty, further investigations of `optimized scale
settings', as performed at N$^3$LO in ref.~\cite{Zeng:2015gha} are not
warranted.

%
\section{Higgs decay to bottom quarks}
\label{sec:hbb}
\setcounter{equation}{0}
%

We denote the perturbative expansion of the function $\widetilde{R}(q^2)$
in eq.~(\ref{GamHbb}) by 
\beq
\label{RtilExp}
  \frac{1}{N_R}\,\widetilde{R}(q^2) \;=\; 
   1 \,+ \sum_{n=1}\:\!\tilde{r}_{n\,}^{}\ar(n)(q^2)
\eeq
in terms of the reduced coupling defined in eq.~(\ref{C1exp}). The coefficients
up to order $\ar(4)$ read, for QCD and its generalization to any simple compact
gauge group,
\bea
\label{Hbb1}
  \tilde{r}_1^{} &\!\! =\!\! &
         17\,\* \cf
 \; , \\[2mm]
\label{Hbb2}
  \tilde{r}_2^{} &\!\! =\!\! &
         \cfs \, \*  \Bigg[
            { 691 \over 4 }
          - 36\, \* \z2
          - 36\, \* \z3
          \Bigg]
       + \,\ca \* \cf \, \*  \Bigg[
            { 893 \over 4 }
          - 22\, \* \z2
          - 62\, \* \z3
          \Bigg]
\nn \\ & & \mbox{}
       - \,\cf\, \* \nf \, \*  \Bigg[
            { 65 \over 2 }
          - 4\, \* \z2
          - 8\, \* \z3
          \Bigg]
 \; , \\[2mm]
\label{Hbb3}
  \hspace*{2mm}\tilde{r}_3^{} &\!\! =\!\! &
         \cft \*  \Bigg[
            { 23443 \over 12 }
          - 648\, \* \z2
          - 956\, \* \z3
          + 360\, \* \z5
          \Bigg]
       + \ca \* \cfs  \*  \Bigg[
            { 13153 \over 3 }
          - 1532\, \* \z2
          - 2178\, \* \z3
          + 580\, \* \z5
          \Bigg]
\nn \\ & & \mbox{\hspn}
       + \,\cas \* \cf \, \*  \Bigg[
            { 3894493 \over 972 }
          - { 6860 \over 9 }\, \* \z2
          - { 4658 \over 3 }\, \* \z3
          + { 100 \over 3 }\, \* \z5
          \Bigg]
\nn \\ & & \mbox{\hspn}
       - \,\ca \* \cf\, \* \nf \, \*  \Bigg[
           { 267800 \over 243 }
         - { 2284 \over 9 }\, \* \z2
         - { 704 \over 3 }\, \* \z3
         + { 48 \over 5 }\, \* \zss
         - { 80 \over 3 }\, \* \z5
          \Bigg]
\\ & & \mbox{\hspn}
       - \,\cfs\, \* \nf \, \*  \Bigg[
           { 2816 \over 3 }
         - 260\, \* \z2
         - 520\, \* \z3
         - { 48 \over 5 }\, \* \zss
         + 160\, \* \z5
          \Bigg]
       \,+ \,\cf \* \nfs \, \*  \Bigg[
            { 15511 \over 243 }
          - { 176 \over 9 }\, \* \z2
          - 16\, \* \z3
          \Bigg]
\nn \; , \\[3mm]
\label{Hbb4}
  \hspace*{2mm}\tilde{r}_4^{} &\!\! =\!\! &
         \cff \* \Bigg[
            { 915881 \over 48 }
          - 8388\, \* \z2
          - 15218\, \* \z3
          + 288\, \* \zss
          + 1296\, \* \z2 \* \z3
          + 7770\, \* \z5
          + 768\, \* \zts
          - 1890\, \* \z7
          \Bigg]
\nn \\ & & \mbox{\hspn}
       + \,\ca \* \cft \, \*  \Bigg[
            { 11553691 \over 144 }
          - { 70445 \over 2 }\, \* \z2
          - { 331541 \over 6 }\, \* \z3
          + { 7602 \over 5 }\, \* \zss
          + 6192\, \* \z2 \* \z3
\nn \\ & & \mbox{\hspp}
          + 31975\, \* \z5
          + 3960\, \* \zts
          - { 32949 \over 2 }\, \* \z7
          \Bigg]
\nn \\ & & \mbox{\hspn}
       + \,\cas \* \cfs \, \*  \Bigg[
            { 830983045 \over 7776 }
          - { 953327 \over 18 }\, \* \z2
          - { 450971 \over 6 }\, \* \z3
          + { 7758 \over 5 }\, \* \zss
          + 9724\, \* \z2 \* \z3
\nn \\ & & \mbox{\hspp}
          + 21955\, \* \z5
          + { 880 \over 7 }\, \* \zst
          + 856\, \* \zts
          + { 41517 \over 4 }\, \* \z7
          \Bigg]
\nn \\ & & \mbox{\hspn}
       + \,\cat \* \cf \, \*  \Bigg[
            { 2087145095 \over 23328 }
          - { 672739 \over 27 }\, \* \z2
          - { 238519 \over 6 }\, \* \z3
          + { 1739 \over 3 }\, \* \zss
          + { 15004 \over 3 }\, \* \z2 \* \z3
\nn \\ & & \mbox{\hspp}
          - { 93875 \over 9 }\, \* \z5
          - { 880 \over 7 }\, \* \zst
          + 5976\, \* \zts
          - { 10899 \over 4 }\, \* \z7
          \Bigg]
\nn \\ & & \mbox{\hspn}
       - \,\dfFAnr \, \*  \Bigg[
           144
         - 1300\, \* \z3
         - 72\, \* \zss
         + 2440\, \* \z5
         - 4896\, \* \zts
         + 1680\, \* \z7
          \Bigg]
\nn \\ & & \mbox{\hspn}
       + \,\cft\, \* \nf \, \*  \Bigg[
          - { 151297 \over 9 }
          + 6889\, \* \z2
          + { 46399 \over 3 }\, \* \z3
          - { 66 \over 5 }\, \* \zss
          - 1584\, \* \z2 \* \z3
\nn \\ & & \mbox{\hspp}
          - 9380\, \* \z5
          - { 480 \over 7 }\, \* \zst
          - 1368\, \* \zts
          + 3360\, \* \z7
          \Bigg]
\nn \\ & & \mbox{\hspn}
       - \,\ca \* \cfs\, \* \nf \, \*  \Bigg[
           { 83380613 \over 1944 }
         - { 162944 \over 9 }\, \* \z2
         - { 79736 \over 3 }\, \* \z3
         + { 3444 \over 5 }\, \* \zss
         + 3128\, \* \z2 \* \z3
\nn \\ & & \mbox{\hspp}
          + { 23314 \over 3 }\, \* \z5
          - { 80 \over 7 }\, \* \zst
          + 652\, \* \zts
          + 1680\, \* \z7
          \Bigg]
\nn \\ & & \mbox{\hspn}
       - \,\cas \* \cf\, \* \nf \, \*  \Bigg[
           { 72695765 \over 1944 }
         - 11949\, \* \z2
         - { 158515 \over 18 }\, \* \z3
         + { 2123 \over 5 }\, \* \zss
         + 1936\, \* \z2 \* \z3
\nn \\ & & \mbox{\hspp}
          - { 65812 \over 9 }\, \* \z5
          - { 400 \over 7 }\, \* \zst
          + 700\, \* \zts
          - 280\, \* \z7
          \Bigg]
\nn \\ & & \mbox{\hspn}
       + \,\dfFFnr\, \* \nf \, \*  \Bigg[
            348
          - 2008\, \* \z3
          - 144\, \* \zss
          - 1152\, \* \zts
          + 2560\, \* \z5
          \Bigg]
\nn \\[1mm] & & \mbox{\hspn}
       + \,\cfs\, \* \nfs \, \*  \Bigg[
            { 7009861 \over 1944 }
          - { 13210 \over 9 }\, \* \z2
          - { 8146 \over 3 }\, \* \z3
          + { 204 \over 5 }\, \* \zss
          + 352\, \* \z2 \* \z3
          + 192\, \* \zts
          + { 2848 \over 3 }\, \* \z5
          \Bigg]
\nn \\[1mm] & & \mbox{\hspn}
       + \,\ca \* \cf \* \nfs \*  \Bigg[
            { 18248293 \over 3888 }
          - { 16031 \over 9 }\, \* \z2
          - { 4972 \over 9 }\, \* \z3
          + { 324 \over 5 }\, \* \zss
          + 304\, \* \z2 \* \z3
          - 32\, \* \zts
          - { 10484 \over 9 }\, \* \z5
          \Bigg]
\nn \\[1mm] & & \mbox{\hspn}
       - \,\cf\, \* \nft \, \*  \Bigg[
           { 520771 \over 2916 }
         - { 2200 \over 27 }\, \* \z2
         - { 260 \over 9 }\, \* \z3
         + { 8 \over 3 }\, \* \zss
         + { 64 \over 3 }\, \* \z2 \* \z3
         - { 160 \over 3 }\, \* \z5
          \Bigg]
\; .
\eea
Eqs.~(\ref{Hbb1}) -- (\ref{Hbb3}) agree with the literature, see 
refs.~\cite{Gorishnii:1990zu,HbbN3LO} and references therein.
The N$^4$LO coefficient $\tilde{r}_4^{}$ has been computed in 
ref.~\cite{HbbN4LO} for the case of QCD. Accordingly setting $\ca = 3$, 
$\cf = 4/3$, $d_F^{\,abcd}d_A^{\,abcd}/N_{\!R} = 5/2$ and
$d_F^{\,abcd}d_F^{\,abcd}/N_{\!R} = 5/36$ in eq.~(\ref{Hbb4}), we find 
complete agreement with their result.

Only the case of $\nf = 5$ is phenomenologically relevant, yet it is
instructive to briefly~consider the numerical dependence of $\widetilde{R}$ 
on the number of light flavours $\nf$ over a wide range,
\bea
 \nf=1  &\!:&
  1
  \,+\, 1.8037560\,\als  \,+\, 3.5038193\,\as(2)
  \,+\, 4.470933\,\as(3) \,-\, 1.765010 \,\as(4)
  \,+\, \ldots
\; , \;\quad
\nn \\
 \nf=2  &\!:& 
  1
  \,+\, 1.8037560\,\als  \,+\, 3.3661592\,\as(2)
  \,+\, 3.664830\,\as(3) \,-\, 3.736264 \,\as(4)
  \,+\, \ldots
\; , \;\quad
\nn \\
 \nf=3  &\!:& 
  1
  \,+\, 1.8037560\,\als  \,+\, 3.2284991\,\as(2)
  \,+\, 2.875431\,\as(3) \,-\, 5.511190 \,\as(4)
  \,+\, \ldots
\; , \;\quad
\nn \\
 \nf=4  &\!:& 
  1 
  \,+\, 1.8037560\,\als  \,+\, 3.0908390\,\as(2)
  \,+\, 2.102737\,\as(3) \,-\, 7.091048 \,\as(4) 
  \,+\, \ldots 
\; , \;\quad
\nn \\
\label{HbbNumNf}
 \nf=5  &\!:&
  1
  \,+\, 1.8037560\,\als  \,+\, 2.9531789\,\as(2)
  \,+\, 1.346747\,\as(3) \,-\, 8.477010 \,\as(4)
  \,+\, \ldots
\; , \;\quad
\\
 \nf=6  &\!:& 
  1 
  \,+\, 1.8037560\,\als  \,+\, 2.8155188\,\as(2)
  \,+\, 0.607462\,\as(3) \,-\, 9.670604 \,\as(4) 
  \,+\, \ldots 
\nn \; . \;\quad
\eea
The main trend is similar to that of the larger coefficients for 
$\Gamma_{H\ra\,gg}$ in eqs.~(\ref{KSInumNf}) and (\ref{KOSnumNf}): 
the $\nf$-dependent coefficients decrease with increasing $\nf$. 
The main difference is that the fourth-order term is already negative 
at $\nf=1$. The $\as(3)\:\!$-term changes sign close to $\nf=7$, 
leading to the large fourth-order$\,/\,$third-order ratio at $\nf=5$ 
already observed in ref.~\cite{HbbN4LO}.

\pagebreak

The break-up of the coefficients $\tilde{r}$ for QCD into `genuine' and 
`kinematic' contributions can be found in eq.~(7) of ref.~\cite{HbbN4LO}. 
The numerical scale dependence of $\widetilde{R}$ has been included in the 
comprehensive study of Higgs decays to hadrons to order $\as(4)$ in 
ref.~\cite{DaviesSW17}. However, the scale dependence of the coefficients
$\rs{n}$ is available in the literature only to order $\as(3)$ \cite{HbbN3LO}.

For the convenience of the reader, we therefore conclude our brief account
of $\Gamma_{H \ra\, \bar{b}^{}b}$ by writing down the generalization of the 
coefficients (\ref{Hbb1}) -- (\ref{Hbb4}) to a general scale $\mu^2\,$:
\bea
  \rs1(L_q) &\! =\! &
       \rs1
       \,-\, 2\*\gam0 \*\, L_q
 \:\: , \nn \\[0.5mm]
  \rs2(L_q) &\! =\! &
       \rs2{}
       \,-\, (\, 2\*\gam1 + 2\*\rs1\*\gam1 + \rs1\*\b0 \,) \* L_q
       \,+\, (\, 2\*\Gam(0,2) + \b0 \* \gam0 \,) \* \, L_q^2
 \:\: , \nn \\[1mm]
  \rs3(L_q) &\! =\! &
       \rs3
       \,-\, (\, 2\*\gam2 + 2\*\rs2\*\gam0 + 2\*\rs2\*\b0 + 2\*\rs1\*\gam1 
              + \rs1\*\b1 \,) \* L_q
\nn \\[0.5mm] & & \quad\mbox{}
       \,+\, (\, 4\*\gam0\*\gam1 + \b1\*\gam0 + 2\*\b0\*\gam1 
             + 2\*\rs1\*\Gam(0,2) + 3\*\rs1\*\b0\*\gam0 + \rs1\*\B(0,2) \,) 
             \* L_q^2
\nn \\[0.5mm] & & \quad\mbox{}
       \,-\, 1/3\,\* (\,4\*\Gam(0,3) + 6\*\b0\*\Gam(0,2) + 2\*\B(0,2)\*\gam0
             \, )
             \*\, L_q^3
 \:\: , \nn \\[1mm]
\label{Hqq4mu}
  \rs4(L_q) &\! =\! &
       \rs4
       \,-\, (\, 2\*\gam3 + 2\*\rs3\*\gam0 + 3\*\rs3\*\b0 + 2\*\rs2\*\gam1 
             + 2\*\rs2\*\b1 + 2\*\rs1\*\gam2 + \rs1\*\b2 \,) \* L_q
\nn \\[0.5mm] & & \quad\mbox{}
       \,+\, (\, 2\*\Gam(1,2) + 4\*\gam0\*\gam2 + \b2\*\gam0 + 2\*\b1\*\gam1 
             + 3\*\b0\*\gam2 + 2\*\rs2\*\Gam(0,2) + 5\*\rs2\*\b0\*\gam0
\nn \\ & & \qquad\quad\mbox{}
             + 3\*\rs2\*\B(0,2) + 4\*\rs1\*\gam0\*\gam1 + 3\*\rs1\*\b1\*\gam0 
             + 4\*\rs1\*\b0\*\gam1 + 5/2\,\*\rs1\*\b0\*\b1 \,) L_q^2
\nn \\[0.5mm] & & \quad\mbox{}
       \,-\, 1/3\,\* (\, 12\*\Gam(0,2)\*\gam1 + 6\*\b1\*\Gam(0,2) 
             + 18\*\b0\*\gam0\*\gam1 + 5\*\b0\*\b1\*\gam0 + 6\*\B(0,2)\*\gam1
\nn \\ & & \qquad\quad\mbox{}
             + 4\*\rs1\*\Gam(0,3) + 12\*\rs1\*\b0\*\Gam(0,2) 
             + 11\*\rs1\*\B(0,2)\*\gam0 + 3\*\rs1\*\B(0,3) \,) L_q^3
\nn \\[0.5mm] & & \quad\mbox{}
       \,+\, 1/6\, \* (\, 4\*\Gam(0,4) + 12\*\b0\*\Gam(0,3) 
             + 11\*\B(0,2)\*\Gam(0,2) + 3\*\B(0,3)\*\gam0 \,) L_q^4
\; ,
\eea
in terms of $L_q\,=\,\ln (q^2/\mu^2)$, $\rs{n}$ 
in eqs.~(\ref{Hbb1}) - (\ref{Hbb2}), the coefficients $\b{n}$ of 
the beta function, and the coefficients $\gam{n}$ of the mass 
anomalous dimension in the \MSb\ scheme up to N$^3$LO, see refs.~%
\cite{Chetyrkin:1997dh,Vermaseren:1997fq} and references therein.
The coefficients to $\rs{3}(L_q)$ agree with eq.~(17) of ref.~\cite{HbbN3LO}.


%
\section{The electromagnetic R-ratio}
\label{sec:rratio}
\setcounter{equation}{0}

The non-singlet and singlet contributions to the electromagnetic R-ratio,
$r(q^2)$ and $r_{\rm S}^{}(q^2)$ in eq.~(\ref{RratEM}), can be expanded 
in the same manner as $N_R^{-1}\widetilde{R}(q^2)$ in eq.~(\ref{RtilExp}). 
At $\mu^2 = q^2$ the coefficients for the dominant non-singlet part read 
\bea
\label{R1}
  r_1^{} &\!\! =\!\! &
         3\,\* \cf
 \; , \\[2mm]
\label{R2}
  r_2^{} &\!\! =\!\! & \mbox{}
       - { 3 \over 2 }\, \* \cfs 
       \,+\, \ca \* \cf \, \*  \Bigg[
            { 123 \over 2 }
          - 44\, \* \z3
          \Bigg]
       \,-\, \cf \* \nf \, \*  \Big[
            11
          - 8\, \* \z3
          \Big]
 \; , \\[2mm]
\label{Rns3}
  \hspace*{2mm}r_3^{} &\!\! =\!\! &
       - { 69 \over 2 }\, \* \cft 
       \,- \,\ca \* \cfs \, \*  \Big[
            127
          + 572\, \* \z3
          - 880\, \* \z5
          \Big]
\nn \\[1mm] & & \mbox{\hspn}
       + \,\cas \* \cf \, \*  \Bigg[
            { 90445 \over 54 }
          - { 242 \over 3 }\, \* \z2
          - { 10948 \over 9 }\, \* \z3
          - { 440 \over 3 }\, \* \z5
          \Bigg]
       \,-\,\cfs\, \* \nf \, \*  \Bigg[
            { 29 \over 2 }
          - 152\, \* \z3
          + 160\, \* \z5
          \Bigg]
\nn \\ & & \mbox{\hspn}
       - \,\ca \* \cf\, \* \nf \, \*  \Bigg[
            { 15520 \over 27 }
          - { 88 \over 3 }\, \* \z2
          - { 3584 \over 9 }\, \* \z3
          - { 80 \over 3 }\, \* \z5
          \Bigg]
       \,+\,\cf\, \* \nfs \, \*  \Bigg[
            { 1208 \over 27 }
          - { 8 \over 3 }\, \* \z2
          - { 304 \over 9 }\, \* \z3
          \Bigg]
\; , \qquad \\[2mm]
\label{Rns4}
  \hspace*{2mm}r_4^{} &\!\! =\!\! &
         \cff \, \*  \Bigg[
            { 4157 \over 8 }
          + 96\, \* \z3
          \Bigg]
       \,-\, \ca \* \cft \, \*  \Big[
           2024
         + 278\, \* \z3
         - 18040\, \* \z5
         + 18480\, \* \z7
          \Big]
\nn \\ & & \mbox{\hspn}
       - \,\cas \* \cfs \, \*  \Bigg[
           { 592141 \over 72 }
         - 121\, \* \z2
         + { 87850 \over 3 }\, \* \z3
         - { 104080 \over 3 }\, \* \z5
         - 9240\, \* \z7
          \Bigg]
\nn \\ & & \mbox{\hspn}
       + \,\cat \* \cf \, \*  \Bigg[
            { 52207039 \over 972 }
          - { 16753 \over 3 }\, \* \z2
          - { 912446 \over 27 }\, \* \z3
          + { 10648 \over 3 }\, \* \z2\, \* \z3
          - { 155990 \over 9 }\, \* \z5
\nn \\[-0.1mm] & & \mbox{\hspp\hspp}
          + 4840\, \* \zts
          - 1540\, \* \z7
          \Bigg]
\nn \\ & & \mbox{\hspn}
       + \,\dfFAnr \, \*  \Big[
            48
          - 64\, \* \z3
          - 320\, \* \z5
          \Big]
       \,-\, \nf\, \* \dfFFnr \, \*  \Bigg[
           208
         + 256\, \* \z3
         - 640\, \* \z5
          \Bigg]
\nn \\ & & \mbox{\hspn}
       + \cft\, \* \nf \, \*  \Bigg[
            { 1001 \over 3 }
          + 396\, \* \z3
          - 4000\, \* \z5
          + 3360\, \* \z7
          \Bigg]
\nn \\ & & \mbox{\hspn}
       + \,\ca \* \cfs\, \* \nf \, \*  \Bigg[
            { 32357 \over 108 }
          + 66\, \* \z2
          + { 42644 \over 3 }\, \* \z3
          - { 41240 \over 3 }\, \* \z5
          - 1056\, \* \zts
          - 1680\, \* \z7
          \Bigg]
\nn \\ & & \mbox{\hspn}
       - \,\cas \* \cf\, \* \nf \, \*  \Bigg[
           { 4379861 \over 162 }
         - 2988\, \* \z2
         - { 137744 \over 9 }\, \* \z3
         + 1936\, \* \z2\, \* \z3
         - { 75220 \over 9 }\, \* \z5
         + 704\, \* \zts
         - 280\, \* \z7
          \Bigg]
\nn \\ & & \mbox{\hspn}
       + \,\cfs \* \nfs \, \*  \Bigg[
            { 5713 \over 27 }
          - 16\, \* \z2
          - { 4648 \over 3 }\, \* \z3
          + { 4000 \over 3 }\, \* \z5
          + 192\, \* \zts
          \Bigg]
\nn \\ & & \mbox{\hspn}
       + \,\ca \* \cf\, \* \nfs \, \*  \Bigg[
            { 340843 \over 81 }
          - 520\, \* \z2
          - { 20906 \over 9 }\, \* \z3
          + 352\, \* \z2\, \* \z3
          - { 10880 \over 9 }\, \* \z5
          - 32\, \* \zts
          \Bigg]
\nn \\ & & \mbox{\hspn}
       - \,\cf\, \* \nft \, \*  \Bigg[
           { 49048 \over 243 }
         - { 88 \over 3 }\, \* \z2
         - { 3248 \over 27 }\, \* \z3
         + { 64 \over 3 }\, \* \z2\, \* \z3
         - { 160 \over 3 }\, \* \z5
          \Bigg]
\; .
\eea
Additional singlet contributions enter from the third order in $\als$, viz
\bea
\label{Rsg3}
  \hspace*{2mm}r_{3,S}^{} &\!\! =\!\! &
          \dtFFnr\,\, \*  \Bigg[
            { 176 \over 3 }
          - 128\, \* \z3
          \Bigg]
\; , \\[2mm]
\label{Rsg4}
  \hspace*{2mm}r_{4,S}^{} &\!\! =\!\! &
         \dtFFnr\, \* \Bigg( \ca \, \*  \Bigg[
            { 31144 \over 9 }
          - 5408\, \* \z3
          + 2880\, \* \z5
          - 1408\, \* \zts
          \Bigg]
\nn \\ & & \mbox{\hspn}
      - \cf \, \*  \Big[
            832
          + 1024\, \* \z3
          - 2560\, \* \z5
          \Big]
       - \nf \, \*  \Bigg[
            { 4768 \over 9 }
          - 832\, \* \z3
          + 640\, \* \z5
          - 256\, \* \zts
          \Bigg]
          \,\Bigg)
\eea
with $d_F^{\,abc}d_F^{\,abc}/N_R = 5/18$ in QCD; for the `time-dependent' 
normalization of this colour factor see the discussion below eq.~(30) of 
ref.~\cite{MVVgam5}.
The above results are in complete agreement with previous calculations, 
see refs.~\cite{Gorishnii:1990vf,Surguladze:1990tg,Rratio1,Rratio2,%
Rratio3,RratioRC} and references therein. 
The fourth-order coefficients (\ref{Rns4}) and (\ref{Rsg4}) had not been 
verified by a second calculation before.

The numerical expansion of the  non-singlet contribution $r(q^2)$ in QCD
is given by
\bea
\label{RnsNumNf}
  \nf = 1 &\!\! : &
       1 + 0.3183099\,\als   + 0.1895124\,\as(2)
         - 0.252925  \,\as(3) - 1.422960 \,\as(4) + \ldots
\; , \nn \\[0.5mm]
  \nf = 2 &\!\! : &
       1 + 0.3183099\,\als   + 0.1778305\,\as(2)
         - 0.213173 \,\as(3) - 1.253232 \,\as(4) + \ldots
\; , \nn \\[0.5mm]
  \nf = 3 &\!\! : &
       1 + 0.3183099\,\als   + 0.1661486\,\as(2)
         - 0.331673 \,\as(3) - 1.097226 \,\as(4) + \ldots
\; , \nn \\[0.5mm]
  \nf = 4 &\!\! : &
       1 + 0.3183099\,\als   + 0.1544668\,\as(2)
         - 0.371548 \,\as(3) - 0.953617  \,\as(4) + \ldots
\; , \nn \\[0.5mm]
  \nf = 5 &\!\! : &
       1 + 0.3183099\,\als   + 0.1427849\,\as(2)
         - 0.411757 \,\as(3) - 0.821078 \,\as(4) + \ldots
\; , \nn \\[0.5mm]
  \nf = 6 &\!\! : &
       1 + 0.3183099\,\als   + 0.1311030\,\as(2)
         - 0.452301 \,\as(3) - 0.698289 \,\as(4) + \ldots
\;\quad .
\eea
The physically relevant numbers of effectively massless flavours are 
$\nf = 3,\,\ldots,6.$ The overall effect of the quartic group invariants 
on the fourth-order coefficient is between 3\% and 5\% for these values 
of $\nf$.
 
The $\als$-corrections in eq.~(\ref{RnsNumNf}) are much smaller than 
their counterparts for $H \rightarrow gg$ in eq.~(\ref{HggNumNf}) and 
$H \rightarrow \bar{b}^{}b$ in eq.~(\ref{HbbNumNf}); as in the latter 
case the $\nf$-dependent coefficients decrease with increasing $\nf$.
The fourth-order correction is largest for low values of $\nf\,$: 
$r_4$ amounts to 5.6 times $r_3$ at $\nf=1$. 
For three flavours the $\as(4)$ correction contributes as much as the 
$\as(3)$ terms at $\als \simeq 0.3$.
This situation is at least exacerbated by the kinematic $\pi^2$ 
contributions, as can be seen from the example decompositions
\bea
\label{RnsNumNfz2}
  \nf = 1 &\!\! : &
  \ldots + ( 0.4551676 - \underline{0.7080921}^{} ) \,\as(3) 
         + ( 1.0596193 - \underline{2.4825797}^{} ) \,\as(4)
  + \ldots
\nn \; , \nn \\[0.5mm]
  \nf = 3 &\!\! : &
  \ldots + ( 0.2054750 - \underline{0.5371479}^{} ) \,\as(3) 
         + ( 0.5038103 - \underline{1.6010363}^{} ) \,\as(4)
  + \ldots
\eea
where, as above, those contributions have been underlined.
For the full $\nf$-dependence of this decomposition see eq.~(7) of
ref.~\cite{Rratio1}.

The generalization of the coefficients in eqs.~(\ref{R1}) -- (\ref{Rns4})
to $\mu^2 \neq q^2$ can be obtained from eqs.~(\ref{Hqq4mu}) by dropping
the terms with $\gam{n}$ which arise from the Yukawa-coupling prefactor
$m_b^2 \equiv m_b^2(\mu^2)$ in eq.~(\ref{GamHbb}). 
The resulting numerical dependence of $r(q^2)$ is very small at the
particularly important point $q^2 = \MZs$, see figs.~2 and 3 of 
ref.~\cite{Rratio3}.

The transformation of eqs.~(\ref{RnsNumNf}) to the miniMOM scheme yields,
with $\als = \alpha_{\rm s,\,MM}$ (here),
\bea
\label{RnsNumNfMM}
  \nf = 1 &\!\! : &
       1 + 0.3183099\,\als   - 0.1390779\,\as(2)
         - 0.780651 \,\as(3) - 0.511193 \,\as(4) + \ldots
\; , \nn \\[0.5mm]
  \nf = 2 &\!\! : &
       1 + 0.3183099\,\als   - 0.1226150\,\as(2)
         - 0.736947 \,\as(3) - 0.342317 \,\as(4) + \ldots
\; , \nn \\[0.5mm]
  \nf = 3 &\!\! : &
       1 + 0.3183099\,\als   - 0.1061521\,\as(2)
         - 0.692733 \,\as(3) - 0.190425 \,\as(4) + \ldots
\; , \nn \\[0.5mm]
  \nf = 4 &\!\! : &
       1 + 0.3183099\,\als   - 0.0896891\,\as(2)
         - 0.648007 \,\as(3) - 0.054783 \,\as(4) + \ldots
\; , \nn \\[0.5mm]
  \nf = 5 &\!\! : &
       1 + 0.3183099\,\als   - 0.0732262\,\as(2)
         - 0.602769 \,\as(3) + 0.065345 \,\as(4) + \ldots
\; , \nn \\[0.5mm]
  \nf = 6 &\!\! : &
       1 + 0.3183099\,\als   - 0.0567633\,\as(2)
         - 0.557020 \,\as(3) + 0.170696 \,\as(4) + \ldots
\; , \quad 
\eea
in agreement with the corresponding parts of eq.~(3.5) -- (3.10) in 
ref.~\cite{RratioJAG} (see also Ref.~\cite{Kataev:2015yha}), where the 
expansion has been written down in terms of $\ars = \als/(4\:\!\pi)$. 
The qualitative pattern in eq.~(\ref{RnsNumNfMM}) is rather different from 
that in eq.~(\ref{RnsNumNf}): here the ratios of the third-order and 
second-order coefficients are large. 
If the fourth-order results were not known, one might by tempted to expect 
a further rapid growth of the coefficients at this order. Yet, the actual 
numbers are much smaller than their third-order counterparts for the physical 
values of $\nf$.

\begin{figure}[p]
\vspace{-2mm}
\centerline{\epsfig{file=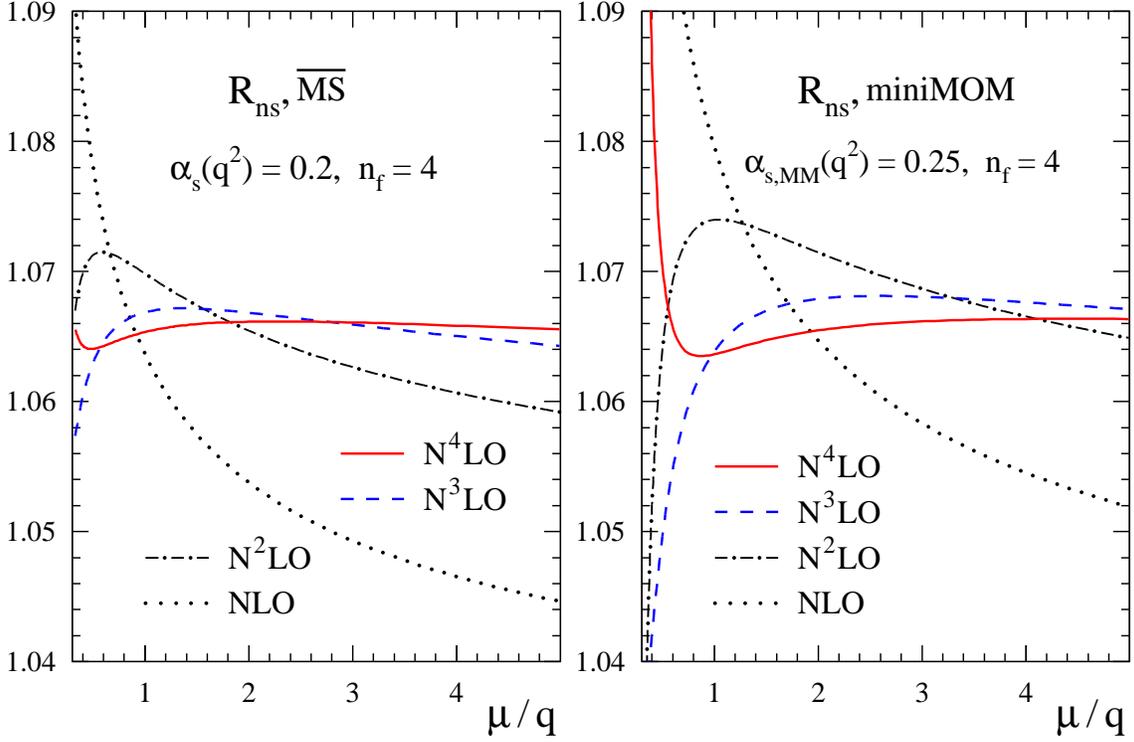,width=15.5cm,angle=0}}
\vspace{-2mm}
\caption{ \label{fig:rscale4} \small
 The renormalization scale dependence of the non-singlet $R$-ratio for
 $\nf=4$ at a reference scale, specified by $\als(q^2) = 0.2$ in \MSb, 
 below the $\Upsilon$ threshold in the \MSb\ and miniMOM schemes.
 }
\vspace*{-2mm}
\end{figure}
\begin{figure}[p]
\vspace{-1mm}
\centerline{\epsfig{file=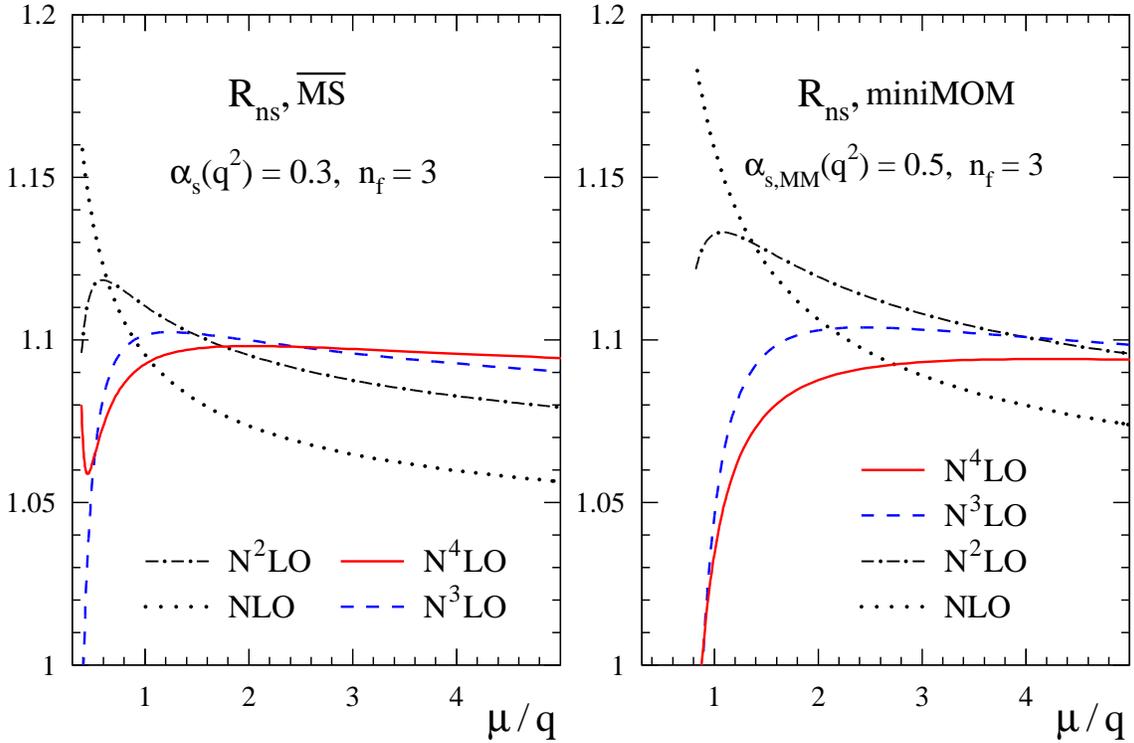,width=15.5cm,angle=0}}
\vspace{-1mm}
\caption{ \label{fig:rscale3} \small
 As fig.~\ref{fig:rscale4}, but for a scale below the $J/\psi$ threshold
 with $\als = 0.3$ in \MSb\ and $\nf = 3$. The~curves have been cut off at 
 low scales where the respective values of $\als$ at N$^3$LO exceed 0.7.} 
\end{figure}

The generalization of eq.~(\ref{RnsNumNfMM}) to $\mu^2 \neq q^2$ is
again given by eqs.~(\ref{Hqq4mu}) with $\gam{n}=0$, but with the \MSb\
beta function replaced by its miniMOM counterpart \cite{MiniMOM1,MiniMOM2}.
The $\mu$-dependence of the $R$-ratio up to order $\as(4)$ is shown for
both schemes in figs.~\ref{fig:rscale4} and \ref{fig:rscale3} at two
low-$q^2$ reference points.
The first is above the $\bar{c}^{}c$ resonances but below the $\Upsilon$
threshold, where an analysis with $\nf = 4$ is appropriate \cite{Rratio3}.
The second is below the $J/\psi$ resonance with $\nf =3$.

The respective scales are specified via (order-independent) \MSb\ values
of $\als(q^2)$, the corresponding miniMOM values of $\als$ are rounded 
results of the N$^4$LO conversion of ref.~\cite{N4LOmMOM}. 
The \MSb\ scale variation in fig.~\ref{fig:rscale3} amounts to 
$R_{\rm ns\!}-\!1 \,=\, (6.51 \pm 0.11)\cdot 10^{-2}$ at N$^4$LO. 
The corresponding miniMOM band is consistent with this result, and only 
slightly wider if the small-$\mu$ spike is not taken into account.
For a very low $q^2$ with $\als(q^2)=0.3$, the results become unstable
below about $\mu = q$ in \MSb\ and $\mu = 2\!\: q$ in miniMOM. 
Disregarding these regions, the N$^4$LO results are fairly stable with a 
3\% uncertainty and $R_{\rm ns}-\!1 \,=\, (9.5 \pm 0.3)\cdot 10^{-2}$ 
in the \MSb\ scheme.


%
\medskip
\section{Summary}
\label{sec:summary}
%

We have completed the N$^4$LO corrections,
i.e., the contributions of order $\as(6)$, for the decay of the Higgs boson
to hadrons via $H \ra\, gg$ at the leading order in the heavy-top limit.
This correction is slightly smaller than the $1/m_{\rm top}^{}$ effects at 
NNLO \cite{Schreck:2007um} but, in all likelihood, larger than the 
presently unknown $1/m_{\rm top}$ correction at N$^3$LO. 
Hence our result provides an improvement of the overall accuracy of 
$\Gamma_{H \ra\, gg}$. The remaining uncertainty due to the truncation
of the perturbation series can be estimated, rather conservatively, as
$\pm 0.6\%$. This is much smaller than the uncertainty of 2.5\% induced 
by a 1\% uncertainty of $\als(\MZs)$.
An~experimental uncertainty of 1\% is, of course, not feasible at the LHC.
However, a future linear $e^+e^-$ collider may be able to address the
coupling of the Higgs boson to gluons at this level, see section 2.3 of
ref.~\cite{Fujii:2015jha}.

Furthermore we have calculated, also for a general gauge group, the 
fourth-order corrections to $H \ra\, \bar{b}^{}b$ in the massless limit 
and to the electromagnetic $R$-ratio for $e^+ e^- \ra \mbox{ hadrons}$. 
These corrections have been computed by one group before in 
refs.~\cite{HbbN4LO} (where the result is presented only for the gauge 
group SU(3)) and \cite{Rratio1,Rratio2,Rratio3,RratioRC}, respectively; 
we find complete agreement with those results. 

Our calculations have been performed using the {\sc Forcer} program
\cite{FORCER} and, at five loops, a {\sc Form} implementation of the 
$R^*$-methods introduced very recently in ref.~\cite{HerzogRuijl17}.
These methods differ substantially from the global $R^*$-method used in
refs.~\cite{HbbN4LO,Rratio1,Rratio2,Rratio3,RratioRC,beta4a}.
Up to four loops we were easily able to keep all powers of the gauge
parameter. This was prohibitively expensive at the five-loop level where we
worked in the Feynman gauge. 
These results have been checked in two ways. The first is verifying the 
correctness of the higher poles in $\ep$, which have to cancel against the 
effect of lower-order diagrams in the renormalization procedure. 
The second check is the computation of all five-loop diagrams in at least 
two different ways.

We have illustrated the size and renormalization-scale dependence of
$\Gamma_{H \ra\, gg}$ and the \mbox{$R$-ratio} in both the standard \MSb\ 
scheme and the miniMOM scheme. The N$^4$LO results and their stability in 
these two schemes are comparable for $\Gamma_{H \ra\, gg}$ and $R$-ratio 
at high scales~$q^{2\,}$; the miniMOM results for the $R$-ratio at 
renormalization scale $\mu^2 \simeq q^2$ become unstable at higher $q^2$ 
than their \MSb\ counterparts.
Overall the miniMOM scheme does not appear to be preferable over \MSb\
in cases, such as the ones considered here, where the perturbation series 
is known to a high order.

{\sc Form} files with our results can be obtained from the preprint server 
http://arXiv.org~by downloading the source of this article. 
They are also available from the authors upon request.

%
\medskip
\subsection*{Acknowledgements}
This work has been supported by the {\it European Research Council}$\,$ (ERC) 
Advanced Grant 320651, {\it HEPGAME} and the UK {\it Science \& Technology
Facilities Council}$\,$ (STFC) grant ST/L000431/1. 
We also are grateful for the opportunity to use most of the {\tt ulgqcd} 
computer cluster in Liverpool which was funded by the STFC grant ST/H008837/1.
The Feynman-diagram figures were made using {\tt Axodraw2} \cite{Axodraw2}. 

%

{
\setlength{\baselineskip}{0.535cm}

}
\end{document}